%% file: NCTIT4-14.tex
\documentclass[times, onecolumn,11pt]{IEEEtran}
\usepackage{amssymb}
\usepackage{tipa}
\usepackage{pifont}
\usepackage{amssymb}
\usepackage{stmaryrd}
 \usepackage{amsmath}

\usepackage{subfigure}
\usepackage{epsfig}
\usepackage{graphicx}
\usepackage{setspace}
\usepackage{colortbl}
\newcommand {\Graph} {{\mathcal G}}
\newcommand {\Edges} {{\cal E}}

\newcommand {\Nodes} {{\mathcal V}}
\newcommand {\Capacity} {C}
\newcommand {\Field} {\mathbb F}

\newcommand {\IRV}[1]{{\mathbf{t'}(#1)}}
\newcommand {\VIRV}[1]{{\mathbf{t''}(#1)}}

\newcommand {\IRM}{T'}
\newcommand {\VIRM}{T''}
\newcommand {\GEV}[1]{\mathbf{t(#1)}}
\newcommand {\GEM}{T}
\newcommand {\Errors} {Z}
\newcommand {\ErrorsMat} {E}
\newcommand {\bl} {n}

\newcommand {\Tran} {T}

\newcommand {\ProEdgFail}{p}

\newcommand {\ProRandomNonSpan}{p_s}
\newcommand {\ProEdgeCorr}{p_c}
\newcommand {\ProNonAcp}{p_a}

\newcommand {\ErrEdg} {{\cal Z}}
\newcommand {\SizErr} {z}

\newcommand {\Path} {\mathcal P}

\newcommand {\Incoming} {\textbf{In}}
\newcommand {\Outgoing} {\textbf{Out}}

\newcommand {\eX} {X}
\newcommand {\eY} {Y}
\newcommand {\RevMat} {Y}

\newcommand{\IRVCand} {\mathcal I_{IRV}}
\newcommand {\FindNode} {\bar{\Nodes}}
\newcommand {\FindEdge} {\bar{\Edges}}
\newcommand {\FindIRV}{{\bar{\cal I}_{IRV}}}
\newcommand {\FindIrv}[1]{{\mathbf{\bar{t}'(#1)}}}
\newcommand {\FindGraph}{\bar{\Graph}}
\newcommand {\RkErr}{{\eta}}

\newcommand {\EdgInj} {{\bf z}}
\newcommand {\ExtEdgesSetOut} {Ext}
\newcommand {\ID}[1]{{{id(#1)}}}

\newtheorem{theorem}{Theorem}
\newtheorem{corollary}[theorem]{Corollary}

\newtheorem{lemma}[theorem]{Lemma}

\begin{document}

\title{Passive network tomography for erroneous networks: A network coding approach}

\author{Hongyi Yao$^{1}$,  Sidharth Jaggi$^{2}$ and Minghua Chen$^{2}$\\
$^{1}$ Tsinghua University $\;$ $^{2}$ The Chinese University of
Hong Kong $\;$
\thanks{This work was supported in part by National Natural Science Foundation of
China Grant 60553001, the National Basic Research Program of China
Grant 2007CB807900 and 2007CB807901, RGC GRF grant 412608, 411008,
and 411209, RGC AoE grant on Institute of Network Coding
established under the University Grant Committee of Hong Kong, CUHK
MoE-Microsoft Key Laboratory of Human-centric Computing and
Interface Technologies, Direct Grant (Project Number 2050397) of The
Chinese University of Hong Kong, and two gift grants from Microsoft
and Cisco. Preliminary versions of this paper are partly in~\cite{NCTInfocom} and \cite{NRSC2010}.
} }

\maketitle

\begin{abstract}
Passive network tomography uses end-to-end observations of network communication to characterize the network, for instance to estimate the network topology and to localize random or adversarial glitches. Under the setting of linear network coding this work provides a comprehensive study of passive network tomography in the presence of network (random or adversarial) glitches. To be concrete, this work is developed along two directions: 1. Tomographic upper and lower bounds (i.e., the most adverse conditions in each problem setting under which network tomography is possible, and corresponding schemes (computationally efficient, if possible) that achieve this performance) are presented for random linear network coding (RLNC). We consider RLNC designed with common randomness, i.e., the receiver knows the random code-books all nodes. (To justify this, we show an upper bound for the problem of topology estimation in networks using RLNC without common randomness.) In this setting we present the first set of algorithms that characterize the network topology exactly. Our algorithm for topology estimation with random network errors has time complexity that is polynomial in network parameters. For the problem of network error localization given the topology information, we present the first computationally tractable algorithm to localize random errors, and prove it is computationally intractable to localize adversarial errors. 2. New network coding schemes are designed that improve the tomographic performance of RLNC while maintaining the desirable low-complexity, throughput-optimal, distributed linear network coding properties of RLNC.  In particular, we design network codes based on Reed-Solomon codes so that a maximal number of adversarial errors can be localized in a computationally efficient manner even without the information of network topology.
The tomography schemes proposed in the paper can be used to monitor networks with other glitches such as packets losses and link delays, etc.

\end{abstract}

\input{Intro.tex}

\input{ProblemFormulation.tex}

\input{MathPreliminaries.tex}

\input{TopoforRLNC.tex}

\input{LocateErrforRLNC.tex}

\input{NetworkRSCode.tex}

\section{Conclusion and Future Work}
This work examines passive network tomography on networks performing
linear network coding in the presence of network errors. We consider both
random and adversarial errors. In part I, under random linear
network coding (RLNC) we give characterizations of when it is
possible to find the topology, and thence the locations of the
network errors. Under RLNC, many of the algorithms we provide have
polynomial time computational complexity in the network size; for
those that are not efficient, we prove intractability by showing
reductions to computationally hard problems. In part II, we design
network Reed-Solomon coding (NRSC) to address the undesirable
tomography capabilities of RLNC under some (especially adversarial error) settings, and
yet preserving the key advantages of RLNC.

Possible future work can proceed in many directions.
\begin{enumerate}
\item Adversarial nodes cannot be located exactly in
general networks. For instance, when the adversarial node $u$
pretends it is receiving erroneous transmissions from its upstream neighbor $v$, it
is impossible for the receiver  to determine whether the error is located at $u$
or at $v$. Hence tomography schemes that {\it approximately} locate adversarial nodes
are hoped for. 

\item Tomography schemes that approximately estimate the
network topology in the presence of adversarial errors are hoped for.


\item The question of designing a network coding scheme that enables efficient topology estimation in the presence of adversarial errors, and yet preserves key advantages of RLNC (low-complexity rate-optimal distributed coding), is open.
\end{enumerate}

\bibliography{YaoG}

\bibliographystyle{IEEEtran}

\section{Appendix}
\label{sec:Appendix}

\subsection{Network erasure model}
An erasure on edge $e$ means that the packet ${\mathbf x}(e)$
carried by $e$ is treated as an all-zeroes length-$\bl$ vector over
$\Field_q$ by the node receiving ${\mathbf x}(e)$, {\it i.e.}, the injected erroneous packet ${\bf z}(e)$ equals $-{\bf x}(e)$.
Two network erasure models are considered:

\begin{enumerate}
\item {\it Random erasures}: Every edge $e$ in $\Edges$ experiences random erasures independently.

\item {\it Adversarial erasures}: The edges that suffer erasures are adversarially chosen.
\end{enumerate}

\subsection{Topology estimation for network erasures under RLNC}

Since adversarial erasures is a weaker attack model than adversarial
errors, the results in Section~\ref{subsec:Topo-adv-RLNC} can be
directly applied to the case of adversarial erasures.

The topology estimation scheme for random erasures is slightly
different from that for random errors. The difference comes from the
fact that in the random error model the injected errors in $\Errors$ are
chosen at random, while in the random erasure model the
injected errors are exactly the negative of the messages
transferred. Thus Lemma~\ref{LeRandomErr} for random error model is
not always true for the random erasure model. Hence we need the
Lemma~\ref{LeRandomEra} below as an alternative. 

Let $\ErrEdg$ be the set of edges
suffering erasures and $|\ErrEdg|=\SizErr$.
Let $\GEV{e}\in \Field_q^{1\times C}$ be the global encoding
vectors~\cite{Raymondbook2008}
 of edge $e$, {\it i.e.}, the packet carried by $e$ is $\GEV{e}\eX$
 when no errors or erasures happen in the network. Let
 $\GEM(\ErrEdg)\in \Field_q^{\SizErr\times C}$ be the matrix whose
 rows comprise of $\{\GEV{e}, e\in \ErrEdg\}$. Recall that $\ErrorsMat=\IRM(\ErrEdg)\Errors$ (as defined
 in Equation~(\ref{eq:net_tran_err})), where the rows of $\Errors$ comprise of $\{{\bf z}(e):e\in\ErrEdg\}$,
 {\it i.e.}, $\{-{\bf x}(e):e\in\ErrEdg\}$.  Then we have:
\begin{lemma}
\label{LeRandomEra} If the source has max-flow $\SizErr$ to the
headers of the edges in $\ErrEdg$, with probability at least
$1-{|\Edges|}/{q}$, the matrix $\Errors$ of injected errors  has full row
rank $\SizErr$ and thus ${\bf \ErrorsMat}={\bf \IRM}(\ErrEdg)$. 
\end{lemma}
{\bf Proof:} Since the network is directed and acyclic,  for ease of
analysis we impose an partial order on the edges of
$\ErrEdg=\{e_1,e_2,...,e_\SizErr\}$. In particular, for any $j>i$,
$e_j$ can not be upstream of $e_j$.

Similarly to Lemma~\ref{LeIRVandEdges}, if the source has max-flow
$\SizErr$ to the headers of the edges in $\ErrEdg$, $\GEM(\ErrEdg)$
has full row rank $\SizErr$ with a probability at least $1-|\Edges|/q$
under RLNC.

The error corresponding to the erasure on $e_1$ equals $-\GEV {e_1}
\eX$. The packet traversing $e_2$ may be effected by the first
erasure. Hence the error corresponding to the erasure on $e_2$
equals $-(\GEV {e_{2}}-a_{1,2}\GEV {e_1})\eX=-{\bf \bar t}
(e_2)\eX$, where $a_{1,2}=c_{1,2}$ is the {\it unit effect from
$e_1$ to $e_2$}. In general, the error corresponding to the erasure
on $e_i$ equals
\begin{eqnarray*}
-{\bf \bar t}({e_i}) \eX &=&-(\GEV {e_{i}}-\sum_{j=1,2,...,i-1}c_{j,i}{\bf \bar t} (e_j))\eX\\
&=& -(\GEV {e_{i}}-\sum_{j=1,2,...,i-1}a_{j,i}\GEV {e_{j}})\eX,
\end{eqnarray*}
where $c_{j,i}$ is the unit effect from $e_j$ to $e_i$.

Thus $\Errors=-A\GEM(\ErrEdg)\eX$, where
$A\in\Field_q^{\SizErr\times\SizErr}$ and the $(i,j)$'th element of
$A$ equal $-a(j,i)$ with $j<i$, $0$ if $j>i$, $1$ if $i=j$. Then $A$
is invertible. If $\GEM(\ErrEdg)$ has full row rank $\SizErr$ and
$\eX$ has an invertible $\Capacity\times\Capacity$ sub-matrix (for
instance, the header corresponding to the identity matrix used in RLNC), $\Errors$ has full row rank $\SizErr$. Thus we
have that ${\bf \ErrorsMat}={\bf \IRM(\ErrEdg)}$.\hfill $\Box$

To estimate the topology for random erasures under RLNC we use a
two-stage scheme similar to that in Section~\ref{subsec:PTTopo}. That is,
stage $1$ is used for collecting IRV information from multiple source generations, and stage $2$ is used for constructing the topology by the IRV information
collected in stage $1$.

For stage $1$, recall that the identity matrix $I_\Capacity$ is the
header of the source matrix $\eX(i)$, where $i$ denotes the index of
source generation. Thus the header of $\eY(i)$ is
$\eY(i)_h=T-\IRM(\ErrEdg(i))A(i)\GEM(\ErrEdg(i))$, where $A(i)$ is
defined in the proof of Lemma~\ref{LeRandomEra}. For $i_1\neq i_2$,
the difference of the headers $\eY(i_1)_h-\eY(i_2)_h$ is
$\IRM(\ErrEdg(i_2))A(i_2)\GEM(\ErrEdg(i_2))-\IRM(\ErrEdg(i_1))A(i_1)\GEM(\ErrEdg(i_1))$.
Since both $A(i_1)$ and $A(i_2)$ are invertible matrixes, the column
space of $\eY(i_1)_h-\eY(i_2)_h$ equals ${\bf
\IRM}(\ErrEdg(i_1)\cup\ErrEdg(i_2))$ if
$\GEM(\ErrEdg(i_1)\cup\ErrEdg(i_2))$ has rank
$|\ErrEdg(i_1)|+|\ErrEdg(i_2)|$. Thus, $\eY(i_1)_h-\eY(i_2)_h$ could
replace $\ErrorsMat(i)_r$ in {\bf FIND-IRV} to provide the
information of IRVs. Thus with the same assumptions as those in
Section~\ref{subsec:PTTopo}, we can use {\bf FIND-IRV} to collect
the IRV information ($\ErrorsMat(i)_r$ is replaced by
$\eY(i_1)_h-\eY(i_2)_h$ for a pair $(i_1,i_2)\in [1,t\ldots,]\otimes
[1,\ldots,t]$).

For stage $2$, we can directly use {\bf FIND-TOPO} to recover the
topology of the network.

\subsection{Locating erasures under RLNC} \label{sec:loc-erasures} The
algorithm {\bf LOCATE-RANDOM-RLNC} can be also used for locating network
erasures (both random and adversarial), resulting in polynomial-time
algorithms.

To locate {\it random erasures}, Lemma~\ref{LeRandomEra} proves that
when the source has max-flow $|\ErrEdg|$ to the headers of $\ErrEdg$ who suffer erasures, $rank(\Errors)=\SizErr$ and ${\bf \ErrorsMat}={\bf
\IRM(\ErrEdg)}$. Thus {\bf LOCATE-RANDOM-RLNC} can  be
used to locate erasures in the network, with using
$\ErrorsMat$ in Step B instead of  $\ErrorsMat_r$.

To use the efficient algorithm {\bf LOCATE-RANDOM-RLNC} to locate {\it
adversarial erasures}, by Lemma~\ref{LeRandomEra} it is required
that any node has in-degree at least $\SizErr$.  Otherwise, the high complexity
algorithm {\bf LOCATE-ADVERSARY-RLNC} can be used to find the locations
of the adversarial erasures.

{\bf Remark}: The algorithm for locating erasures can also be used for
locating edges experiencing problematic delays. Let
$\eY_d\in\Field_q^{\Capacity\times \bl}$ be the delayed packet
matrix received by $r$. Then $r$ can locate the delayed edges by
treating $\eY_d$ as the erasure matrix $\ErrorsMat$ and then using
the scheme for locating network erasures.
\end{document}

%% file: Intro.tex
\section{Introduction}

The goal of \emph{passive network tomography} (or \emph{passive
network monitoring}) is to use end-to-end observations of
network communication to infer the network topology, estimate
link statistics such as loss rate and propagation delay, and locate
network failures~\cite{NetworkTomo}.

In networks using {\it linear network coding} each node
outputs linear combinations of received packets; this has been shown
to attain optimal multicast throughput~\cite{Raymond2000}. In
fact, even random linear network codes (where each node independently and
randomly chooses the linear combinations used to generate transmitted packets)
suffice to attain the optimal multicast
throughput~\cite{RandCode0}, \cite{RandCode3}, \cite{RandCode2}. In
addition to their desirable distributed nature, such schemes also
have low design and implementation complexity~\cite{RandCode3},
\cite{RandCode2}.

The main observation driving this work is that the linear transforms
arising from random linear network coding have specific
relationships with the network structure, and these relationships
can significantly aid tomography.
Prior work~\cite{Tracyloss1}\cite{SidNCT} has also observed this relationship.

\begin{figure}[htp]
  \begin{center}
    \subfigure  [Routing Case]{\label{ToyExample-a}\includegraphics[height=27mm,width=62mm]{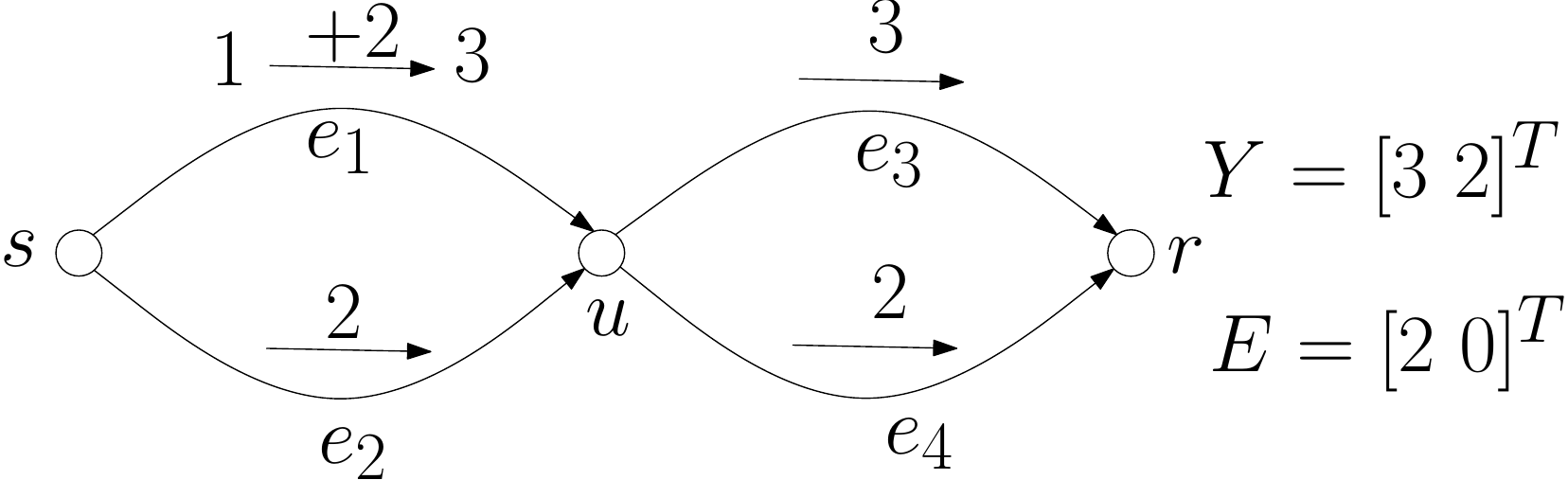}}\label{ToyExample-a}
    \subfigure [Coding Case]{\label{ToyExample-b}\includegraphics[height=27mm,width=62mm]{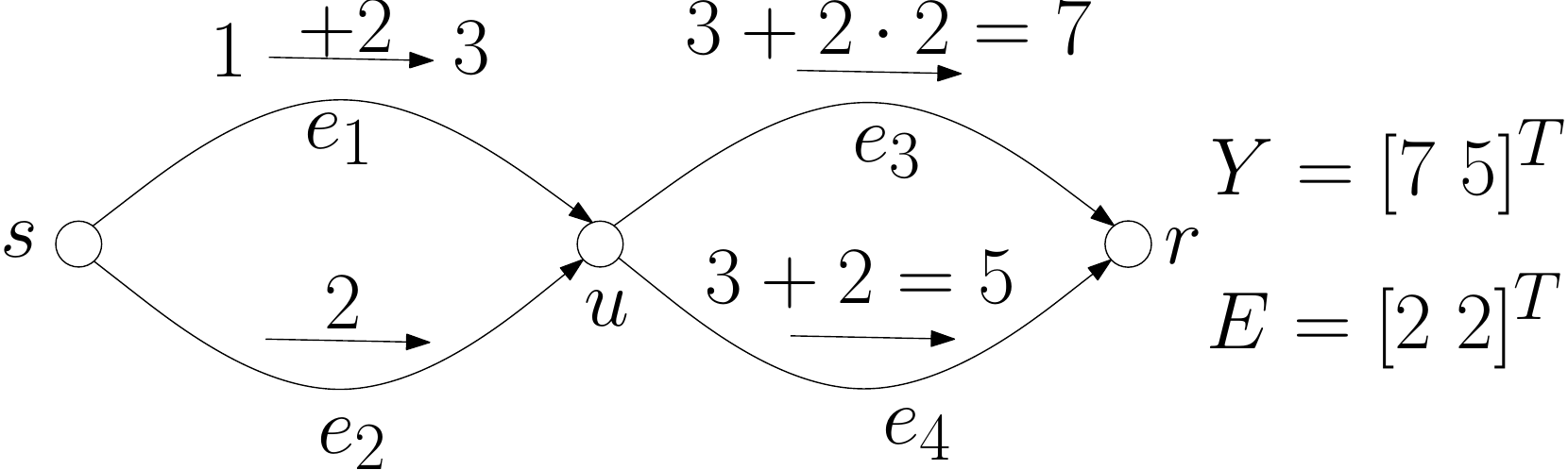}} \label{ToyExample-b}\\
  \end{center}
  \caption{A tomographic example for locating an error at edge $e_1$. In Figure~1(a) observing error vector $\ErrorsMat=[2~0]^T$ is not enough to distinguish the error  locations
$e_1$ and $e_2$. In Figure~1(b), since network coding
is used by intermediate node $u$, the information of $\ErrorsMat=[2~2]^T$ is enough to
locate the erroneous edge $e_1$.}
  \label{ToyExample}
\end{figure}

\noindent {\bf Toy example for error localization:} Consider the tomography problem in
Figure~\ref{ToyExample}. Source $s$ transmits probe symbols (
say $1$ and $2$) to receiver $r$ via intermediate node $u$.
Suppose edge $e_1$ is erroneous and adds (say) $2$ to every symbol
transmitted over it. Receiver $r$ knows the probe symbols, network, and communication schemes {\it a priori}.
It also knows one of the links is erroneous (though it doesn't know in what manner), and wants to locate the erroneous link.

The case where the network communicates only via routing is shown in
Figure~1(a). The probe symbols $1$ and $2$ are
transmitted over edges $e_1$ and $e_2$ respectively to node $u$. Due
to the error introduced over $e_1$, node $u$ receives symbols $3$
and $2$ via edges $e_1$ and $e_2$ respectively, and
forwards them to node $r$ via edges via edges $e_3$ and $e_4$ respectively. Node $r$ receives two symbols from $e_3$ and
$e_4$, denoted by the vector $Y=[ 3 ~2]^T$. Since $r$ knows that probe
symbols {\it a priori}, it can compute the error vector to be
$\ErrorsMat=Y- [1~2]^T=[2~0]^T$. Using $\ErrorsMat$ and its knowledge of the routing scheme, node $r$ can infer that the error happened in the routing path $\{e_1,e_3\}$, but can not
figure out whether the error occurred on $e_1$ or $e_3$.

Figure~1(b) shows the case where node $u$ applies
linear network coding to transmit symbols. In particular, node $u$ outputs
${\bf x_3}={\bf x_1}+2{\bf x_2}$ to link $e_3$ and ${\bf x_4}={\bf
x_1}+{\bf x_2}$ to $e_4$, where ${\bf x_1}$ and ${\bf x_2}$ are the
symbols that node $u$ receives from $e_1$ and $e_2$, and ${\bf x_3}$
and ${\bf x_4}$ are the symbols to be sent over $e_3$ and $e_4$. For
a unit additive error ${\bf e}=1$ at $e_1$, $e_2$, $e_3$ or $e_4$,
the receiver $r$ would observe error vectors ${\bf e}[ 1~1]^T$, ${\bf
e}[ 2 ~1]^T$, ${\bf e}[1 ~0]^T$ or ${\bf e}[ 0 ~1]^T$ respectively.
Thus, errors in different links result in observed error vectors
corresponding to vector spaces. Such linear algebraic characteristics of networks
can be exploited to locate the erroneous link. Specifically, if error ${\bf e}=2$ is injected into $e_1$, node $r$ receives ${Y}=[ 7 ~5]^T$. Knowing in advance the probe symbols and node $u$'s coding scheme, the receiver $r$
computes the error vector as $\ErrorsMat=Y-[ 1+2 \cdot 2 ,1+2]^T=[
2 ~2]^T$. Upon observing $\ErrorsMat=[ 2
~2]^T$ and comparing with the set of possible error vectors corresponding to different error locations, $r$ can determine that $e_1$ is the erroneous
link and the error is ${\bf e}=2$. \hfill $\Box$

While the toy example above might give the impression that the coding scheme needs to be carefully designed for the communication problem at hand, our
results in this paper show that in fact random linear coding suffices to result in tomographic schemes that are distributed and have low computational and communication overhead.
Further, if end-to-end network error-correcting codes
(see for instance~\cite{SidByAdversary}\cite{RankMetricRandCode}) are used for the network communication layer,
in addition network tomography can {\it also} be implemented in a ``passive" manner,
{\it i.e.}, no dedicated probe messages are necessary. Thus
throughout this work, the phrase ``network tomography" stands for
``passive network tomography'' unless otherwise specified.

In this work we consider a network
in which all nodes perform linear network coding.
Besides receiving the messages, the receiver(s) wants to recover the
network topology, and then detect and locate
adversarial attacks, and random glitches (errors or erasures).

We perform a comprehensive study of passive network tomography in
the presence of network errors, under the setting of network coding.
In particular, we seek answers to the following questions:
\begin{itemize}
\item In networks performing random linear network coding (RLNC), what are the appropriate
tomographic upper and lower bounds? That is, what are the most adverse conditions in each problem setting under which network tomography is possible, and what schemes (computationally efficient, if possible) achieve this performance?
\item Are there any linear network coding schemes that improve upon the tomographic upper bounds for RLNC while maintaining their desirable low-complexity, throughput-optimal, distributed linear network coding properties?
\end{itemize}

\subsection{Main contributions}

We now examine the relationship that
linear
transforms arising from random linear network coding have with the structure of the network.
For this we find it useful to
define the \emph{ impulse response vector} (IRV) $\IRV e$ for every
link $e$ as the transform vector from link $e$ to the receiver (see
Section~\ref{subsec:LinTran} for details). As shown in subsequent sections, each $\IRV e$ can be
treated as the fingerprint of corresponding link $e$. Any error on $e$ exposes its
fingerprint, allowing us to detect the location of the error. Note that all the tomography schemes proposed in the paper for network errors can be used to monitor networks with other glitches
such as network erasures ({\it i.e.}, packet losses) and link
delays. We delay discussion on these related topics to
the Appendix.

\begin{itemize}
\item For network tomography under RLNC, our results are categorized into
two classes:

\noindent 1) {\it Topology estimation}. For networks suffering from
random or adversarial errors, we provide the first algorithms (under some sufficient conditions) that estimate the network
topology (in the case of random errors, our algorithms are computationally efficient). We also provide necessary conditions for such topology estimation to be possible (there is currently a gap between our necessary and sufficient conditions).
Common randomness is assumed, {\it i.e.,} that the coding
coefficients of each node are chosen from a random code-book known
by the receiver. Note that the adversaries are allowed to access such knowledge. Without such knowledge, we prove that in the
presence of adversarial or random errors it is either theoretically
impossible or computationally intractable to estimate topology
accurately.

\noindent 2) {\it Error localization}. We provide the first
polynomial time algorithm for locating edges experiencing random
errors. For networks suffering from adversarial errors we provide an upper bound of the number of locatable errors, and also a corresponding (exponential-time) algorithm that matches this bound.
Moreover, we provide the first proof of computational intractability of the
problem. Note that as with error-localization schemes in the previous
literature
(\cite{Tracyloss1,FrauglilossrateAllerton2005,FragouliTopoAllerton2006,FrauglilossrateGlobecom2007}),
the schemes we provide for RLNC require the information of network
topology and the local linear coding coefficients -- this can be from the topology estimation algorithms in this work, or
as part of the network design {\it a priori}.



\item In the other direction, to circumvent the provable tomographic limitations of RLNC, we propose a specific class of random linear network codes that we call network Reed-Solomon coding (NRSC), which have the following three desirable
features:

\noindent 1) NRSC are linear network codes that are implemented in a
distributed manner (each network node only needs to know the node-IDs of its
adjacent neighbors).

\noindent 2) With high probability over code design NRSC achieves the the multicast capacity.

\noindent 3) NRSC aids tomography in the following two aspects:
\begin{itemize}
\item {\it Computational efficiency}. Under the adversarial error model, the receiver can
locate a number of adversarial errors that match a corresponding tomographic upper bound in a computationally
efficient manner.For the random error model, an lightweight topology algorithm is provided under
NRSC.
\item {\it Robustness for dynamic networks}. For  adversarial (and random) error localization the algorithms under NRSC do
 not require the priori knowledge of the network topology and thus are robust against dynamic network updating. For topology estimation in the random error model, the the algorithm under NRSC
 fits for dynamic networks better than the one under RLNC.
\end{itemize}
\end{itemize}
In Table~\ref{tab:SumResult} we compare our results and previous works on computational complexity.

\begin{table}[htb!]
\caption{Comparison our results and previous works on computational complexity } 
\centering 
\begin{tabular}{c|c|c|c|c}
\hline\hline 
Objective &Failure model&Tomography for&Tomography for &Tomography for \\
&&RLNC[Previous works]&RLNC[This work]&NRSC[This work]
\\ [0.5ex]
\hline
&Adversarial Errors & - &Exponential&-\\
\raisebox{1.5ex}{Topology Estimation}& Random Errors&- &Polynomial&Polynomial\\
\hline
&Adversarial Errors &Exponential\cite{Tracyloss1}&Hardness Proof&Polynomial\\
\raisebox{1.5ex}{Failure Localization}&Random Errors&Exponential\cite{Tracyloss1,FrauglilossrateAllerton2005}&Polynomial&Polynomial\\
\hline
\label{tab:SumResult}
\end{tabular}
\end{table}
\subsection{Related work}

\noindent {\it Common randomness:} Essentially all prior tomography results for RLNC assume some form of common randomness, {\it
i.e.,}, the receiver is assumed to have prior knowledge of the random coding
coefficients used by internal nodes. Some previous
results~\cite{SidNCT,Tracyloss1,FrauglilossrateGlobecom2007} for
locating errors under RLNC do not explicitly assume common
randomness, but assume the receiver knows all the linear coding coefficients employed by each node in the network, which is related to our notion of common randomness.

We summarize related work on network inference under the following
categories.
\begin{enumerate}
\item {\it Passive tomography}:
The work in~\cite{SidNCT} provided the first explicit (exponential-time) algorithm for estimating the topology of networks performing RLNC with no errors.
The work in~\cite{Tracyloss1}
studied the problem of locating network errors for RLNC with prior knowledge
of network topology. In particular, error localization can be done in time
${\cal O}{|\Edges|\choose \SizErr}$, where $|\Edges|$ is the number of
links in the network and $\SizErr$ is the number of
errors the network experiences.

\item {\it {\it Active} tomography:} The authors
in~\cite{FrauglilossrateAllerton2005},~\cite{FragouliTopoAllerton2006},
and~\cite{FrauglilossrateGlobecom2007} perform network tomography by using
probe packets and exploiting the linear algebraic structure of network coding. The setting considered in these works concern active tomography, whereas in this work we focus on passive tomography.

\begin{enumerate}
\item {\it Random error localization:} The authors in~\cite{FrauglilossrateAllerton2005}
and~\cite{FrauglilossrateGlobecom2007} study error\footnote{In
fact network erasures are considered in their works. Here we
classify network erasures as a subclass of network errors.}
localization in a network using {\it binary XOR} coding. Using {\it
pre-designed} network coding and probe packets, they show that the
sources can use fewer probe packets than traditional tomography
schemes based on routing. Again, ${\cal O}{|\Edges|\choose \SizErr}$ is the
computational complexity of localization.

\item {\it Topology estimation:} For binary-tree networks using pre-designed
binary XOR coding~\cite{FragouliTopoAllerton2006} show
that the topology can be recovered by using probe packets. The
authors in~\cite{FragouliTopo09} generalize the results to
multi-source multi-receiver scenario.
\end{enumerate}

\item {\it Network inference with internal nodes' information:}
Another interesting set of works
(\cite{FraugliBottleINM2007},~\cite{FraugliSubSpaceITW2007},~\cite{FraugliSubSpaceNetCod2008})
infer the network by the ``packet information" of each internal
node. In particular, the work in~\cite{FraugliSubSpaceITW2007}
discusses the subspace properties of packets received by
internal nodes, the work in~\cite{FraugliBottleINM2007} infers
the bottlenecks of P2P networks using network coding, and the works
in~\cite{FraugliSubSpaceNetCod2008,INFOCOMLocateAdvNode} provides efficient schemes to
locate the adversaries in the networks. Note that  these works require the topology estimator to
have access to internal network nodes.
\end{enumerate}

\subsection{Organization of the paper}
The rest of this paper is organized as follows. We formulate the
problem in Section~\ref{sec:formulation} and present preliminaries
in Section~\ref{sec:prelimilary}. We then present our main technical
results. Our results for network tomography consist
of two parts: Part I considers RLNC, the schemes for topology
estimation in the presence of network adversary and random errors
are presented in Section~\ref{sec:Topo-RLNC}, and the schemes for
error localization is presented in Section~\ref{sec:loc-err-RLNC};
Part II, consists of a particular type of RLNC, network Reed-Solomon
coding (NRSC), in Sections~\ref{Sec:N-RSC}, Section~\ref{sec:locate-adv-rs} and Section~\ref{sec:Topo-Rand-NRSC}.

%% file: ProblemFormulation.tex
\section{Problem Formulation and Preliminaries}
\label{sec:formulation}

\subsection{Notational convention}

Scalars are in lower-case ({\it e.g.} $\SizErr$). Matrices are in
upper-case ({\it e.g.} $\eX$). Vectors are in lower-case bold-face
({\it e.g.} ${\bf e}$). Column spaces of a matrix are in upper-case
bold-face ({\it e.g.} ${\bf E}$). Sets are in upper-case calligraphy
({\it e.g.} ${\mathcal Z}$).

\subsection{Network setting}
\label{subsec:NetSet} For ease of discussion, we consider an direct acyclic
and delay-free network $\Graph=(\Nodes,\Edges)$ where $\Nodes$ is
the set of nodes and $\Edges$ is the set of edges. Each node has a
{\it unique identification number} known to itself. Such a label
could correspond to the node's GPS coordinates, or its IP address,
or a factory stamp. The capacity of each edge is normalized to be
one symbol of $\Field_q$ per unit time. We denote $e(u,v)$ as the
edge from node $u$ to $v$.  For each node $v\in \Nodes$, let
$\Incoming(v)$ be the set of all incoming edges (or nodes) of $v$
and $\Outgoing(v)$ be the set of all outgoing edges (or nodes) of
$v$. The out-degree of node $v$ is defined as $|\Outgoing(v)|$ and  in-degree of node $v$ is defined as $|\Incoming(v)|$.

Note that all the results in the paper can be generalized to the
scenario where edges with non-unit capacity are allowed. Non-unit
capacity edge is modeled as parallel edges, which can be notated by
somewhat unwieldy notations, say $e(u,v,i)$, which stands for the
$i$'th parallel edge from $u$ to $v$.

We focus on the unicast scenario where a single source $s$
communicates with a single receiver $r$ over the network. In
principle, our results can be generalized to other communication
scenarios where RLNC suffices. For instance, in the networks with multiple receivers, we
assume all incoming edges of the receivers are reconnected to
a virtual receiver who performs network tomography.

Let $\Capacity$ be the {\it min-cut} (or {\it max-flow}) from $s$ to
$r$. Without loss of generality, we assume that both the number of
edges leaving the source $s$ and the number of edges entering the
receiver $r$ equal $\Capacity$. We also assume that for every node
in $\Nodes$, there is at least one path between the node and the
receiver $r$; Otherwise the node does not involve in the
communication and is irrelevant to our study.

\subsection{Dependency}\label{subsec:dependency}

Any set of $\SizErr$ edges $e_1,e_2,...,e_{\SizErr}$ is said to be
{\it flow-independent} if there is a path from the tail of each
to the receiver $r$, and these $\SizErr$ paths are edge-disjoint.
The {\it flow-rank} of an edge-set $\ErrEdg$ equals the max-flow
from the tails of edges in $\ErrEdg$ to the receiver $r$. A
collection of edge-sets $\ErrEdg_1,\ErrEdg_2,...,\ErrEdg_n$ is said
to be {\it flow-independent} if $\mbox{\it
flow-rank}(\cup_{i=1}^n\ErrEdg_i)=\sum_{i=1}^n\mbox{\it
flow-rank}(\ErrEdg_i)$. The flow-rank of an internal node
equals to the flow-rank of its outgoing edges. For the set
$\ErrEdg\subseteq \Edges$ with flow-rank $\SizErr $, the {\it
extended set} (or $\ExtEdgesSetOut(\ErrEdg)$) is the set that is of
flow-rank $\SizErr $, includes $\ErrEdg$ and is of maximum size.
Note that $\ExtEdgesSetOut(\ErrEdg)$ is well-defined and
unique~\cite{GraphTheoryBook}.

\subsection{Network transmission via linear network coding}
\label{subsec:lnc}

 In this paper we consider the linear network coding scheme proposed
in~\cite{Raymond2003}.  Let each packet have $n$ symbols from
$\Field_q$, and each edge have the capacity of transmitting one
packet, {\it i.e.}, a row vector in $\Field_q^{1\times n}$.

{\it Source encoder:} The source $s$ arranges the data into a
$\Capacity \times \bl$ {\it message matrix} $\eX$ over $\Field_q$.
Then on each outgoing edge of $s$ a linear combination over
$\Field_q$ of the rows of $\eX$ is transmitted. $\eX$ contains a
pre-determined ``short" {\it header} (say, the identity matrix in $\Field_q^{C\times C}$) known in advance to both the
source and the receiver, to indicate the linear transform from the
source to the receiver.

{\it Network encoders:} Each internal node similarly takes linear
combinations of the packets on incoming edges to generate the
packets transmitted on outgoing edges. Let ${\mathbf x(e)}$
represent the packet traversing edge $e$. An internal node $v$
generates its outgoing packet ${\mathbf x(e')}$ for edge $e'\in
\Outgoing(v)$ as
\begin{equation}
{\mathbf x(e'}) = \sum_{e \in \Incoming(v)} \beta(e,v,e'){\mathbf
x(e)}, \label{Eq:linear-net-coding}
\end{equation}
where $\beta(e,v,e')$ is the linear coding coefficient from the
packet ${\mathbf x(e)}$ to the packet ${\mathbf x(e')}$ via $v$. As a default let $\beta(u,v,w)=\beta(e,v,e')$, where $e=(u,v)$ and
$e'=(v,w)$.

{\it Receiver decoder:} The receiver $r$ constructs the
$C \times \bl$ matrix $\RevMat$ over $\Field_q$ by
treating the received packets as consecutive length-$\bl$ row
vectors of $\RevMat$. 
The network's internal linear operations induce a linear transform
between $\eX$ and $\RevMat$ as
\begin{equation}
\RevMat = \Tran \eX, \label{eq:net_tran}
\end{equation}
where $\Tran\in \Field_q^{C\times C}$ is the overall transform matrix. The receiver $r$ can
extract $\Tran$ from the packet headers (recall
that internal nodes mix headers in the same way as they mix
messages). Once $\Tran$ is invertible the receiver can decode $\eX$ by
$\eX=\Tran^{-1}\eY$.

\subsection{Network error models}
\label{subsec:errormodel} Networks may experience disruption as a
part of normal operation.  Edge errors are considered in this work
-- node errors may be modeled as errors of its outgoing edges.

Let  ${\mathbf x}(e)\in \Field_q^{1\times n}$ be the input packet of
$e$. For each edge $e\in\Edges$ a length-$\bl$ row-vector ${\bf
z}(e)$ is added into  ${\mathbf x}(e)$. Thus the output packet of $e$
is ${\bf y}(e)={\bf x}(e)$+${\bf z}(e)$. Edge $e$ is said to suffer
an error if and only if ${\bf z(e)}$ is a non-zero vector.

Both adversarial and random errors are considered:
\begin{enumerate}
\item  {\it Random errors}: every edge $e$ in $\Edges$ independently experiences random errors with a non-negative
probability.
A random error on $e$ means that ${\bf z(e)}$ has {\it at least} one
randomly chosen position, say $i$, such that the $i$'th symbol of
${\bf z(e)}$ is chosen from $\Field_q$ uniformly at random.
\footnote{Note the difference of this model from the usual model of
{\em dense random errors} on $\Field_q$~\cite{RandomErrorModel},
wherein ${\bf z(e)}$ is chosen from $\Field_q^\bl$ at random. The
model described in this work is more general in that it can handle
such errors as a special case. However, it can {\it also} handle
what we call ``sparse" errors, wherein only a small fraction of
symbols in ${\bf z(e)}$ are non-zero. Such a sparse error may be a
more natural model of some transmission error scenarios~\cite{ZigzagSigcom08,SymbolErrSigCom08}. They may
also be harder to detect. In our model we consider the worst-case
sparsity of $1$. }

\item {\it Adversarial errors}: The network is said to have $\SizErr$ adversarial errors if 
and only if the adversary can arbitrarily choose a subset of edges $\ErrEdg\subseteq \Edges$ with $|\ErrEdg|=\SizErr$ and the
corresponding erroneous packets $\{{\bf z}(e),e\in \ErrEdg\}$. Note that the adversary is assumed to have
unlimited computational capability and has the access to the information of the source matrix $\eX$, network 
topology, all network coding coefficients and tomography algorithms used by the receiver.  
\end{enumerate}

\subsection{Tomography Goals}
\label{subsec:tomo_goals}

The focus of this work is network passive end-to-end tomography in
the presence of network errors. There are two tomographic goals:
\begin{enumerate}
\item {\it Topology estimation}: The receiver $r$ wishes to correctly identify the network topology upstream of it ({\it i.e.}, the graph $\Graph$).
\item {\it Error location}: The receiver $r$ wishes to identify the locations where errors occur in the network.
\end{enumerate}

{\bf Remark}: In fact, all tomography schemes in the paper can be
generalized in the following manner. Instead of the incoming edges
$\Incoming(r)$ of the receiver $r$, consider any cut $\Edges_C$ of
edges that disconnects source $s$ from receiver $r$. A network
manager that has access to the packets output from $\Edges_C$ can
use the tomography schemes in the paper to estimate the topology of
the upstream network and locate the network errors.

\subsection{Network error-correcting codes}
\label{ErrorCorrectCode} Consider the scenario where a randomly or
maliciously faulty set of edges $\ErrEdg$ of size $\SizErr$ {\it
injects} faulty packets into the network. As
in~\cite{RankMetricRandCode}, the network transform
(\ref{eq:net_tran}) then becomes
\begin{eqnarray}
\RevMat= \Tran\eX + \ErrorsMat.\label{eq:err_mat}
\end{eqnarray}
Note that the $\Capacity \times \bl$ {\it error matrix} $\ErrorsMat$ has rank at most $\SizErr$ (see Section~\ref{sub:IRVandNetError} for
details). 
The goal for the receiver $r$ in the presence of such errors is still to reconstruct the source's
message $\eX$. Note that the loss-rate $2\SizErr/\Capacity$ is
necessary and sufficient for correcting $\SizErr$ adversarial
errors~\cite{RankMetricRandCode,SidByAdversary}, while the loss-rate
$(\SizErr+1)/\Capacity$ is necessary and
sufficient~\cite{SidByAdversary} for correcting $\SizErr$ random
errors.

In this work we use the algorithms of~\cite{RankMetricRandCode} for
adversarial errors, and the algorithms of~\cite{SidByAdversary} for
random errors. All our tomography schemes are based on the correct
using of these network error-correcting codes.

\subsection{Computational hardness of NCPRLC}
\label{subsec:NCPHard} Several theorems we prove regarding the
computational intractability of some tomographic problems utilize
the hardness results of the following well-studied problem.

The Nearest Codeword Problem for Random Linear Codes (NCPRLC) is
defined as follows:
\begin{itemize}
\item {\it NCPRLC: $(H,\SizErr,{\bf e})$}:
Given a {\it parity check matrix} $H$ which is chosen uniformly at
random over $\Field_q^{l_1\times l_2}$ with $l_2>l_1$, a constant
$\SizErr$, and a vector ${\bf e}\in {\bf H}$ which is linear combined from
at most $\SizErr$ columns of $H$. The algorithm is required to output a {\it $\SizErr$-sparse
solution} {\bf b} for $H{\bf b}={\bf e}$, {\it i.e.}, ${\bf e} = H {\bf b}$ and
${\bf b}$ has at most $\SizErr$ nonzero components.
\end{itemize}

Note that NCPRLC is a well known computational hard
problem~\cite{ComlexityLinearDecodeing},~\cite{ComlexityLinearDecodeing},~\cite{VardyMDPHard}.

\subsection{Decoding of Reed-Solomon codes}
\label{subsec:RSC} This section introduces some properties of the
well-studied Reed-Solomon codes (RSCs)~\cite{RS-CODE}, used in
particular for worst-case error-correction for point-to-point
channels. A Reed Solomon code (RSC) is a linear error-correcting
code over a finite field $\Field_q$ defined by its parity check
matrix $H\in \Field_q^{l_1\times l_2}$ with $l_2>l_1$. Here $l_1+1$ is {\it
minimum Hamming distance} of the code, {\it i.e.},
minimum number of nonzero components among the codewords belonging to the
code. In particular, $H$ is formed as
\begin{equation}
H=[{\bf h}_1,{\bf h}_2,...,{\bf h}_{l_2}], \label{Eq:RSParity}
\end{equation}
where ${\bf h}_i=[h_i,(h_i)^2,...,(h_i)^{l_1}]^T\in\Field_q^d$ and
$h_i\neq 0$ for each $i\in[1,l_2]$ and $h_i\neq h_j$ for $i\neq j$.

Given  ${\bf e}$ which is a linear combination of any $\SizErr\leq
(l_1+1)/2$ columns of $H$, the decoding algorithm of RS-CODE, denoted as
$\mbox{{\bf RS-DECODE}}(H,{\bf e})$, outputs a {\it $\SizErr$-sparse
solution} of $H{\bf b}={\bf e}$ with $O(l_2l_1)$ operations over
$\Field_q$ (see~\cite{RS-DECODE}). That is, ${\bf b}\in
\Field_q^{l_2}$ has at most $\SizErr$ non-zero components and  ${\bf
e}=H{\bf b}$. Further more, for any ${\bf b'}\neq {\bf b}$ either
${\bf e}\neq H{\bf b'}$ or ${\bf b'}$ has more than $\SizErr$
non-zero components, {\it i.e.}, $\mathbf b$ is the unique
$\SizErr$-sparse solution of $H{\bf b}={\bf e}$.

%% file: MathPreliminaries.tex
\section{Impulse Response Vector (IRV)}
\label{sec:prelimilary}

In this section, we explain the relationship between the linear
transforms induced by the linear network coding and the network
structures, by introducing the concept of  {\it impulse response
vector} (IRV). The relationship forms the mathematical basis for our
proposed tomography schemes.


\subsection{Definition of Impulse Response Vector (IRV)}
\label{subsec:LinTran}

Corresponding to each edge $e \in \Edges$ we define the
length-$\Capacity$ impulse response vector (IRV) $\IRV{e}\in
\Field_q^C$ as the linear transform from $e$ to the receiver.
In particular, let the source $s$ transmit the {\it all-zeroes
packet} ${\mathbf{0}} \in \Field_q^\bl$ on all outgoing edges, let
edge $e$ inject a packet ${\mathbf{z} (e)} \in \Field_q^\bl$, and
let each internal node perform the linear network coding operation. Then the matrix $\RevMat$ received by the receiver $r$ is
$\RevMat=\IRV{e} \mathbf{z}(e)\in \Field_q^{\Capacity\times \bl}$.
So $\IRV{e}$ can be thought of as a ``unit impulse response" from
$e$ to $r$.

An illustrating example for edge IRVs is in Figure~2 and Figure~\ref{Fig:IRVExampleIRV}, where the
coding coefficients are shown in Figure~2 and the packet length is
assumed to be $1$. In Figure~\ref{Fig:IRVExampleIRV}(a), only $e_4$ has an injected symbol
$1$ and what $r$ receives is $\RevMat=[1,0]^T$, thus the IRV of
$e_4$ is $\IRV{e_4}=[1,0]^T$. For the same reason, the IRVs of $e_5,
e_3,e_2$ and $ e_1$ are computed in Figure~\ref{Fig:IRVExampleIRV}(b)
, Figure~\ref{Fig:IRVExampleIRV}(c), Figure~\ref{Fig:IRVExampleIRV}(d)
 and Figure~\ref{Fig:IRVExampleIRV}(e) respectively.

 For a set of edges $\ErrEdg \subseteq \Edges$ with
$|\ErrEdg|=\SizErr$, the columns of the
${\Capacity \times \SizErr}$ {\it impulse response matrix}
$\IRM(\ErrEdg)$ comprise of the set of vectors $\{\IRV{e} : e\in
\ErrEdg\}$.

All IRVs can be inductively computed. First, the IRV for each edge
incoming to the receiver is set as a distinct unit vector, {\it i.e.}, a distinct column of the $C\times C$
identity matrix. Then for
each edge $e$ incoming to node $v$ with outgoing edges
 $\{e_1,e_2,...,e_{d}\}$ we have
 $$\IRV{e} = \sum_{j=1,2,...,d}\beta(e,v,e_j)\IRV {e_j}.$$

\begin{figure}[htp]
 \begin{center}
 {\includegraphics[width=30mm, height=45mm]{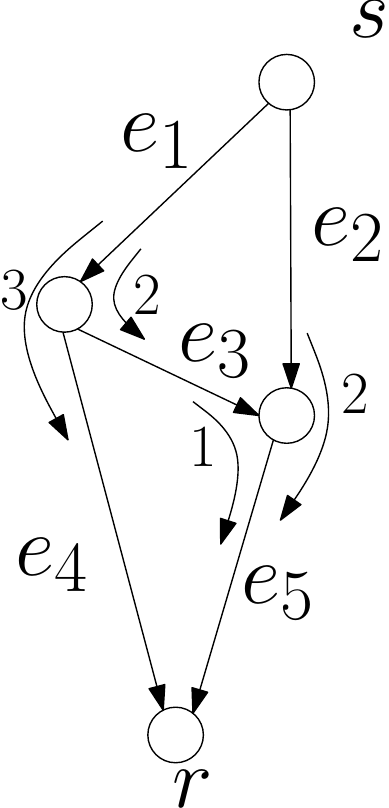}}
\caption{ An example network and its local coding coefficients.}
\end{center}
\label{Fig:IRVToyNet}
\end{figure}

\begin{figure}[htp]
 \begin{center}
 \subfigure[$\IRV{e_4}$]{\includegraphics[height=40mm]{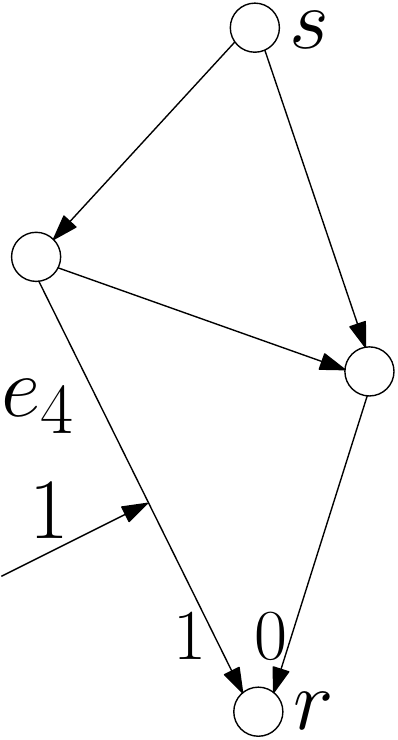}}\label{Fig:IRVExampleIRVB}
 \subfigure[$\IRV{e_5}$]{\includegraphics[height=40mm]{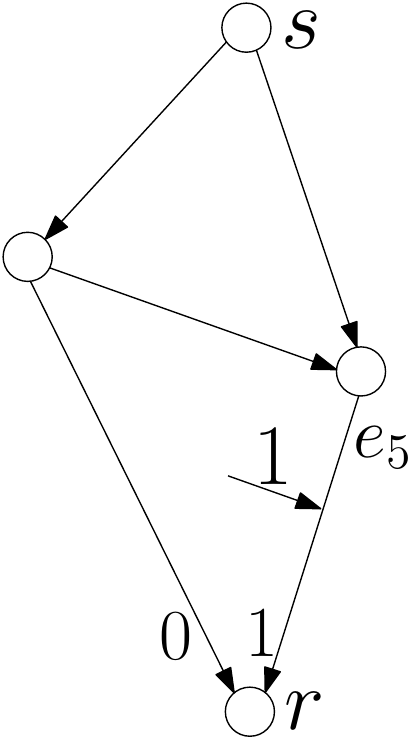}}\label{Fig:IRVExampleIRVC}
 \subfigure[$\IRV{e_3}$]{\includegraphics[height=40mm]{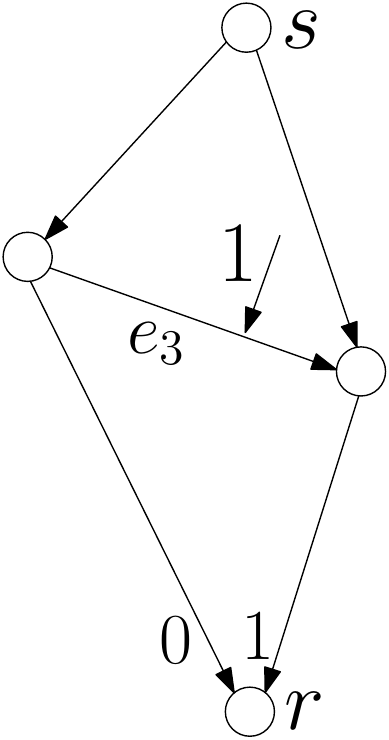}}\label{Fig:IRVExampleIRVD}
 \subfigure[$\IRV{e_2}$]{\includegraphics[height=40mm]{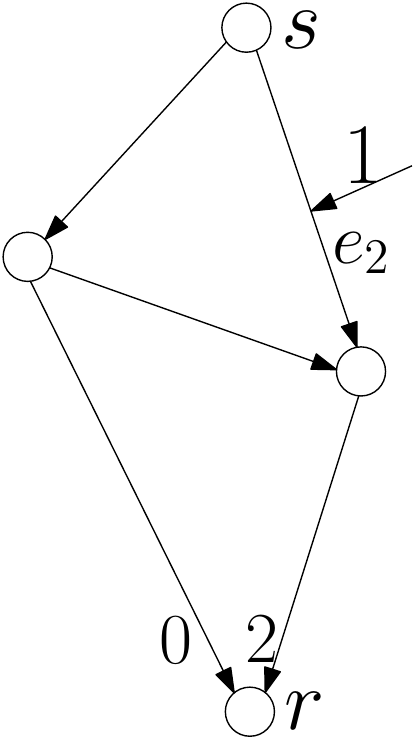}}\label{Fig:IRVExampleIRVE}
 \subfigure[$\IRV{e_1}$]{\includegraphics[height=40mm]{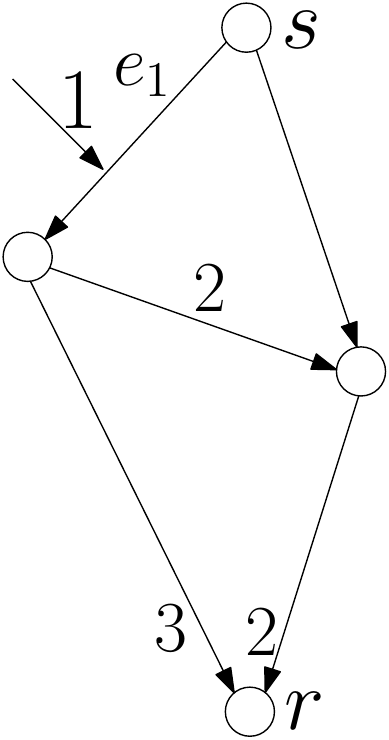}}\label{Fig:IRVExampleIRVF}
 \end{center}
\caption{ The IRVs of the edges shown in Figure~2: $\IRV {e_4}=[1,0]^T$, $\IRV{e_5}=[0,1]^T$,
$\IRV{e_3}=\IRV{e_5}=[0,1]^T$, $\IRV{e_2}=2\IRV{e_5}=[0,2]^T$ and
$\IRV{e_1}=3\IRV{e_4}+2\IRV{e_3}=[3,2]^T$.
 Edges $e_2$ and $e_3$ are not flow-independent, so the IRV $\IRV{e_2}$ equals the
$\IRV{e_3}$ (up to a scalar multiple). Conversely, $e_1$ and $e_5$
are flow-independent, so $\IRV{e_1}$ is linearly independent
from $\IRV{e_5}$.} \label{Fig:IRVExampleIRV}
\end{figure}

\subsection{IRVs under random linear network coding (RLNC)}
\label{subsec:rlnc} The linear network coding defined in
Section~\ref{subsec:lnc} is a random linear network coding (RLNC) if
and only if~\cite{RandCode3}:

 {\it Source encoder:} The source $s$ takes $\Capacity$ independently and uniformly
random linear combinations of the rows of $\eX$ to generate
respectively the packets transmitted on each edge outgoing from $s$
(recall that exactly $\Capacity$ edges leave the source $s$).

{\it Network encoders:} Each internal node, say $v$, independently
and uniformly chooses its local coding coefficients
$\{\beta(e,v,e'),e\in \Incoming(v), e'\in \Outgoing(v)\}$ at random.

{\it Receiver decoder:} As shown in Equation~(\ref{eq:net_tran}), the
receiver $r$ receives $\RevMat$ as $\RevMat = \Tran \eX$ , where
$\Tran$ is the overall transform matrix. It is proved that with a
probability at least $1-|\Edges|/q$ the matrix $\Tran$ is invertible
for RLNC~\cite{RandCode3}. The receiver extracts $\Tran$ from
the header of $Y$ and decodes $\eX$ by
$\eX=\Tran^{-1}\eY$.

For RLNC, the linear transforms defined in Section
\ref{subsec:LinTran} provides an algebraic interpretation for the
graphes. To be concrete, Lemma~\ref{LeIRVandEdges} below states that the linear
independence of the IRVs has a close relationship with
the flow-independence of the edges. The relationship is used in tomography schemes shown in later sections.
\begin{lemma}
\label{LeIRVandEdges}
\begin{enumerate}
\item \label{LeCorIRVEdg} The rank of the impulse response matrix $\IRM(\ErrEdg)$ of an edge set $\ErrEdg$ with flow-rank $\SizErr $ is at most $\SizErr $.

\item \label{LeIndIRVEdg} The IRVs of flow-independent edges are linear independent with a probability at least $1-|\Edges|/q$.

\end{enumerate}
\end{lemma}
\noindent {\bf Proof:}
\begin{enumerate}
\item~\label{pf:1} When the flow-rank of $\ErrEdg$ is $\SizErr $, the max-flow from $\ErrEdg$ to $r$ is
at most $\SizErr $. If the rank of $\IRM(\ErrEdg)$ is larger than
$\SizErr $, say $\SizErr +1$, $\ErrEdg$ can transmit information to
$r$ at rate $\SizErr +1$, which is a contradiction.

\item~\label{pf:3} For an flow-independent edge set $\ErrEdg$ with cardinality $\SizErr $, assume a
virtual source node $s'$ has $\SizErr $ virtual edges connected to
the headers of $\ErrEdg$, and all outgoing edges (except for $\ErrEdg$) of the headers of
$\ErrEdg$ are deleted. The max-flow from $s'$
to $r$ is $\SizErr $ and $\ErrEdg$ is a cut.  Then $\IRM(\ErrEdg)$
has rank $\SizErr $ if and only if $s'$ can transmit information to
$r$ at rate $\SizErr $. By a direct corollary of Theorem 1
in~\cite{RandCode0}, this happens with a probability at least
$1-|\Edges|/q$.

\end{enumerate}
\hfill $\Box$

Thus for a large enough field-size $q$, properties of the edge sets
map to the similar properties of the IRVs. For instance,
with a probability at least $1-|\Edges|/q$, $\mbox{\it
flow-rank}(\cup^d_{i=1}\ErrEdg_i)=\sum^d_{i=1}\mbox{\it
flow-rank}(\ErrEdg_i)$ if and only if
$rank(\IRM(\cup^d_{i=1}\ErrEdg_i))=\sum^d_{i=1}rank(\IRM(\ErrEdg_i))$.
Thus by studying the ranks of $\IRM(\ErrEdg_i)$, we can infer the
flow-rank structures of $\ErrEdg_i$.

The example in Figure 3 also shows the relationship between
flow-independence and linear independence.

\subsection{IRVs for network errors}
\label{sub:IRVandNetError}  Assume a faulty set of edges $\ErrEdg$ of size $\SizErr$ {\it
injects} faulty packets into the network, {\it i.e.}, $\ErrEdg=\{e:e\in \Edges, {\bf z}(e)\neq 0\}$ and $|\ErrEdg|=\SizErr$.
From the definition of IRV, we have:
\begin{eqnarray}
\RevMat = \Tran \eX + \IRM(\ErrEdg)\Errors,\label{eq:net_tran_err}
\end{eqnarray}
where $\Errors$ is a $\SizErr \times \bl$ matrix whose rows comprised of erroneous packets $\{{\bf z}(e): e\in \ErrEdg\}$.
Thus the error matrix $\ErrorsMat$ defined in Equation~(\ref{eq:err_mat}) (of Section~\ref{ErrorCorrectCode}) equals $\IRM(\ErrEdg)\Errors$ and has rank
at most $\SizErr$.

%% file: TopoforRLNC.tex
\begin{center}
 {\large {Part I: Network Tomography for Random Linear Network Coding (RLNC)}}\\
\end{center}

\section{Topology estimation for RLNC}
\label{sec:Topo-RLNC}

\subsection {Common randomness} \label{subsec:SharedRand}
{\it Common randomness} means that all candidate local coding coefficients $\{\beta(u,v,w), u,w\in \Nodes\}$ of node $v\in \Nodes$
 are chosen from its local random
code-book $\mathcal R_v$, and the set of all local random code-books $\mathcal R=\{\mathcal R_v,v\in \Nodes\}$ is known {\it a priori} to the receiver $r$.
Note that $\mathcal R$ can be public to all parties including the adversaries.

 { Common randomness} is
both necessary and sufficient for network topology estimation under
RLNC. On one hand the sufficiency is followed by the works
in~\cite{SidNCT} and this section. On the other
hand we show that in the presence of adversarial (or
random errors), determining the topology without assuming common randomness
is theoretically impossible (or computationally
intractable) later in Theorem~\ref{thm:comm_rand_necc} and
Theorem~\ref{thm:high_comp_rand}.

Each local random code-book in $\mathcal R$ comprises of a list of elements from
$\Field_q$, with each element chosen independently and uniformly at
random. These random code-books can be securely broadcasted by
the source {\it a priori} using a common public key signature scheme
such as RSA~\cite{Rivest78amethod}, or as part of network design.

Depending on the types of failures in the network, we define two
 types of common randomness. Recall
that $\beta(u,v,w)$ is the local coding coefficient from edge $e(u,v)$ via $v$ to the edge $e'(v,w)$ (see
Section~\ref{subsec:lnc} for details).

\begin{enumerate}
\item{\it Weak type common randomness for random errors}:
For node $v\in \Nodes$ each distinct element $(u,w)$ in $\Nodes\otimes \Nodes$ indexes
a distinct element in $\mathcal R_v$. The local
coding coefficient $\beta(u,v,w)$ is chosen as the element $\mathcal
R_v(u,w)$. For instance consider the subnetwork shown in Figure~4. Under
weak type common randomness, Figure~5 shows how node $v_1$ chooses the coding coefficient $\beta(v_2,v_1,v_4)$.

\begin{figure}[htp]
\begin{center}
 \includegraphics[width=40mm]{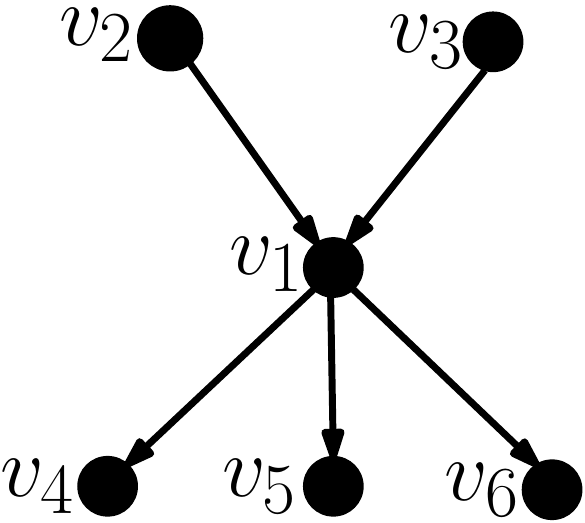}
 \end{center}
 \begin{center}
  \caption{The adjacent neighbors of node $v_1$.}
 \end{center}
 \label{Fig:CommonRandNode}
\end{figure}

\begin{figure}[htp]
\begin{center}
 \includegraphics[width=85mm,height=40mm]{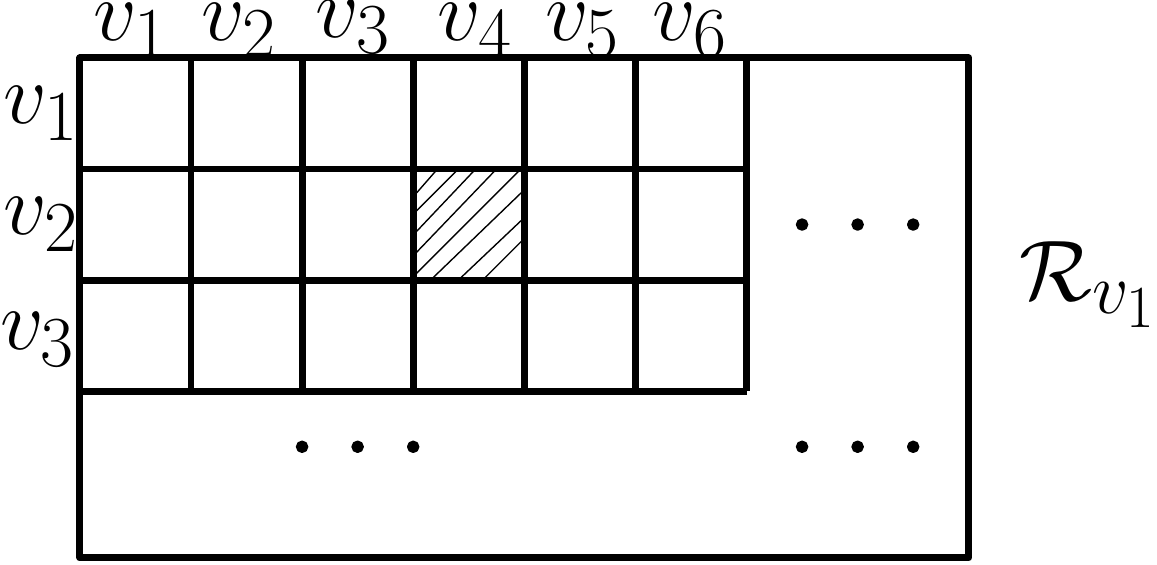}
 \end{center}
 \begin{center}
  \caption{Under weak type common randomness, node $v_1$ in Figure 4 chooses $\beta(v_2,v_1,v_4)$ as the element shown in the dark region.}
 \end{center}
 \label{Fig:CommonRandWeak}
\end{figure}

\item {\it Strong type common randomness for adversarial
errors}:
For node $v\in \Nodes$ each distinct element $(u,w,w')$ in $\Nodes\otimes \Nodes\otimes \Nodes$ indexes
a distinct element in $\mathcal R_v$. For an instance network,
recall that $\Outgoing(v)$ is the outgoing edges of $v$. The coding
coefficients $\beta(u,v,w)$ is chosen as
\begin{equation}\beta(u,v,w)=\sum_{e(v,w')\in
\Outgoing(v)}\mathcal R_v(u,w,w'). \label{eq:TypeII}\end{equation}
For instance consider the subnetwork shown in Figure~4. Under
strong type  common randomness, Figure~6 shows how node $v_1$ chooses the coding coefficient $\beta(v_2,v_1,v_4)$.

\begin{figure}[htp]
\begin{center}
 \includegraphics[width=115mm,height=60mm]{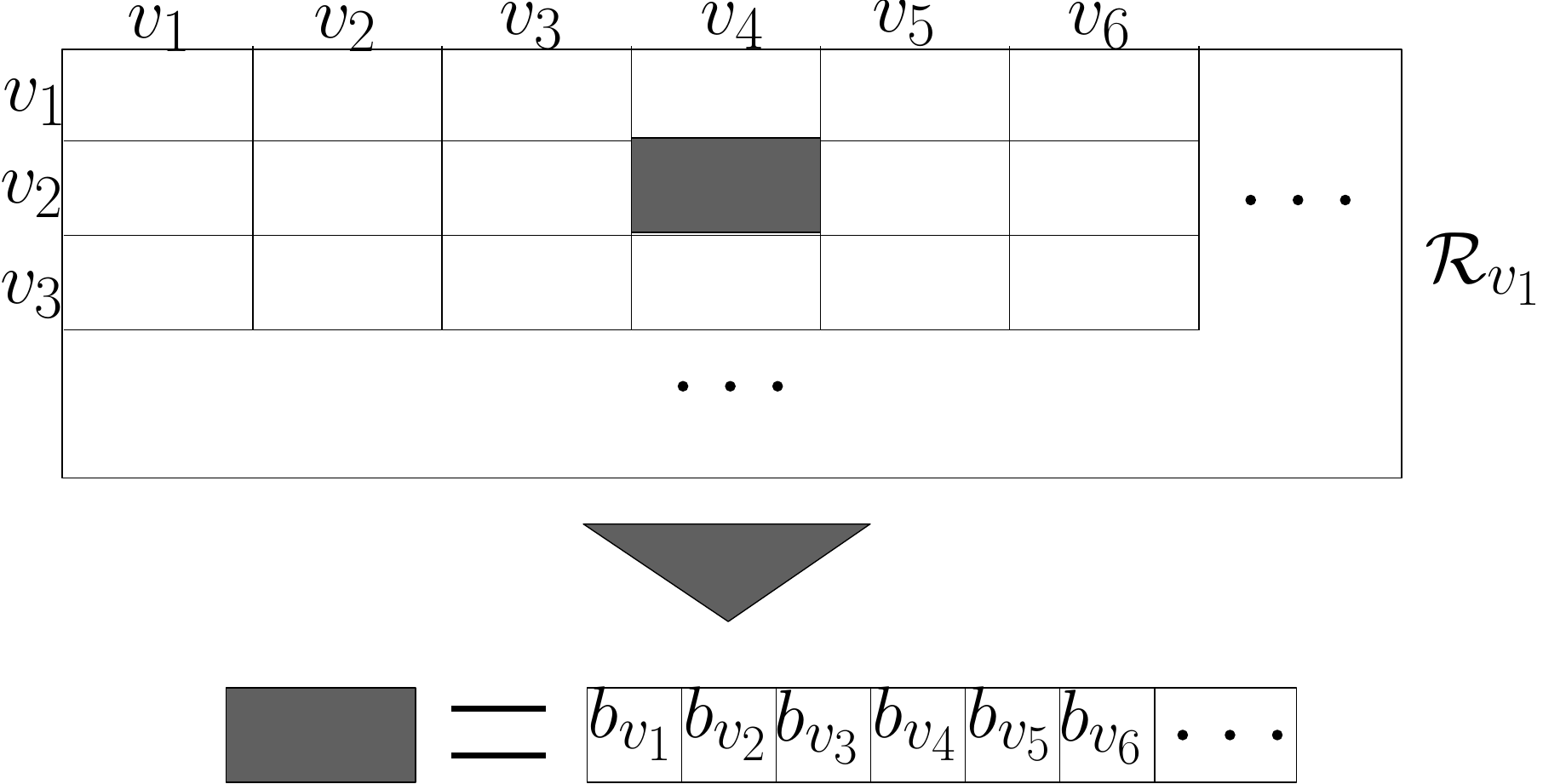}
 \end{center}
 \begin{center}
  \caption{Under strong type common randomness, node $v_1$ in Figure 4 chooses $\beta(v_2,v_1,v_4)$ as $b_{v_4}+b_{v_5}+b_{v_6}$.}
 \end{center}
 \label{Fig:CommonRandStrong}
\end{figure}

\end{enumerate}

\noindent{\bf Remark 1}: For strong type common randomness, since 1) all symbols in $\mathcal R_v$ are independently
and uniformly chosen over finite field $\Field_q$ and 2) for
different coding coefficient $\beta(u,v,w)$ the summation in
equation~(\ref{eq:TypeII}) involves distinct elements in $\mathcal
R_v$, the coding coefficients ${\beta(u,v,w)}$ chosen by equation~(\ref{eq:TypeII}) is also
independently and uniformly distributed over $\Field_q$.

\noindent{\bf Remark 2}: For adversarial errors it is required that the
existence of an edge $e(v,w)$ would effect the coding coefficients
$\{\beta(u,v,w'):w'\neq w\}$. Otherwise, if the adversary corrupts
$e(v,w)$ and only sends all-zero packet on $e(v,w)$, the receiver is
impossible to notice the existence of $e(v,w)$. Thus a different
type of common randomness is used for network
suffering adversarial errors.

\noindent{\bf Remark 3}: Assuming the common randomness, given the knowledge of network topology all local coding
coefficients are known. Thus the IRVs of the edges can be computed efficiently.

\noindent{\bf Remark 4}: For network with parallel edges the random code-book $\mathcal R_v$ can be described by
somewhat unwieldy notations. For instance, under weak common randomness the element $\mathcal R_v(u,w,i,j)$
is for the coding coefficient from edge $(u,v,i)$ ({\it i.e.}, the $i$th parallel edge between $u$ and $v$) to
$(v,w,j)$ via $v$.

We first prove the necessity of using common randomness for
topology estimation in networks with adversarial errors. Since the
network adversaries can hide themselves and only inject zero errors,
it suffices to prove common randomness is necessary for
topology estimation in networks with zero errors.
\begin{theorem}
\label{thm:comm_rand_necc} If internal nodes choose local coding
coefficients independently and randomly {\it without} assuming common randomness, there exist two networks which can not be
distinguished by the receiver in the absence of network errors.
\end{theorem}
\noindent {\bf Proof:} Since the overall transform matrix (see
Equation~(\ref{eq:net_tran}) for details) is the only information the
receiver can retrieve from the receiving packets, it suffices to
prove the overall transform matrixes of $Exp1$ and $Exp2$ in Figure~7 are statistically indistinguishable.

For $Exp1$, let $T_s(1)\in \Field_q^{2\times 2}$ be the transform
matrix from $s$ to $u_1$. 
Similarly, matrices $T_{u_1}\in \Field_q^{3\times 2}$, $T_{u_2}\in
\Field_q^{2\times 3}$, and $T_{u_3}\in \Field_q^{2\times 2}$ are the
transform matrices from $u_1$, $u_2$, $u_3$ to the adjacent
downstream nodes respectively. Thus, the transform matrix $T(1)$
from $s$ to $r$ in $Exp1$ is $T(1)=T_{u_3}T_{u_2}T_{u_1}T_{s}(1)$.

For the similar reason, the transform matrix $T(2)$ from $s$ to $r$
in $Exp2$ is $T(2)=T_{v_3}T_{v_2}T_{v_1}T_{s}(2)$.

Since each element in $T_{u_3}$, $T_{u_2}$, $T_{u_1}$, $T_{s}(1)$,
$T_{v_3}$, $T_{v_2}$, $T_{v_1}$, $T_{s}(2)$ is independently and
uniformly chosen at random, $T(1)$ is  statistically
indistinguishable from $T(2)$. \hfill $\Box$
\begin{figure}[htp]
\begin{center}
 \includegraphics[width=45mm,height=40mm]{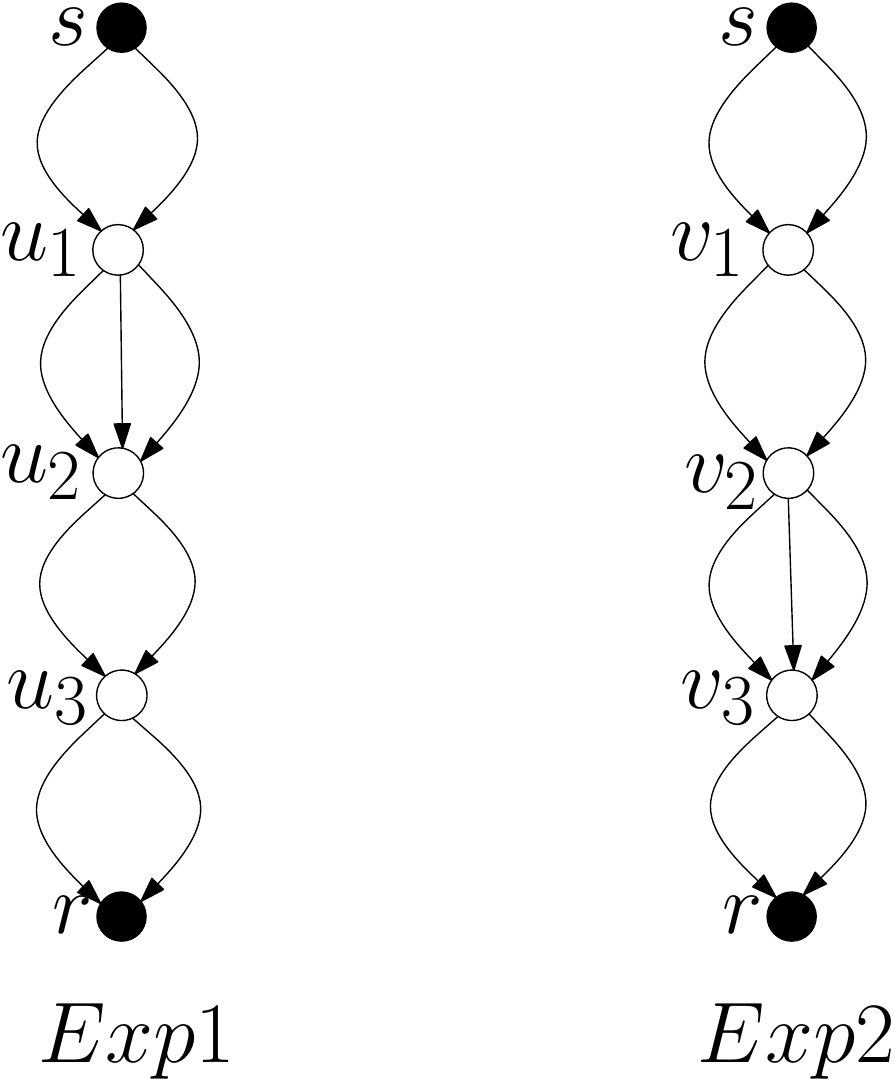}
 \end{center}
 \begin{center}
  \caption{Two networks that are impossible to distinguish by the receiver.}
 \end{center}
 \label{ComplRandImpo}
\end{figure}

For the random error model (see Section~\ref{subsec:errormodel} for
details), Theorem~\ref{thm:comm_rand_necc} does not suffice to show
the necessity of common randomness.
The reason is that in a zero error network the only network
information observed by the receiver is the transform matrix $T$, while
in networks suffering random errors, an random error on the edge may
expose its IRV information which aids topology estimation.  In the
following it is proved that without assuming common randomness
topology estimation is at least as computationally
intractable as NCPRLC (see the definition in
Section~\ref{subsec:NCPHard} for details).

For random error model, as in (\ref{eq:net_tran_err}), the receiver
gets $Y=\Tran\eX+\ErrorsMat$, where
$\ErrorsMat=\IRM(\ErrEdg)\Errors$. Thus ${ \ErrorsMat}$ and $\Tran$
are all  the information observed by the receiver $r$.
Let $\IRVCand$
 be the set of vectors, each of which equals an IRV of an edges in the network.
 Note that $\IRVCand$ is merely a set of vectors,
and as such, individual element has no correspondence with
 any edge in the network.
When the edge suffers random errors independently, $\Errors$
are errors chosen at random. Thus the error matrix
${\ErrorsMat}=\IRM(\ErrEdg)\Errors$ can not provide more information
than $\IRM(\ErrEdg)$, whose columns are in $\IRVCand$. Thus it suffices to prove:

\begin{theorem}
When the internal nodes choose local coding coefficients
independently and randomly {\it without} assuming common randomness, if the receiver $r$ can correctly estimate the topology
in polynomial time (in network parameters) with knowing $\Tran$ and
$\IRVCand$ {\it a priori}, NCPRLC can be solved in time polynomial
(in problem parameters). \label{thm:high_comp_rand}
\end{theorem}
{\bf Proof:} Given a NCPRLC instance $(H,\SizErr,{\bf e})$, as shown
in Figure ~\ref{CommHard}, we construct a network with $l_1$ edges
to $r$ and $l_2$ edges to node $u$.

\begin{figure}
\begin{center}
\label{CommHard}
 \includegraphics[width=30mm]{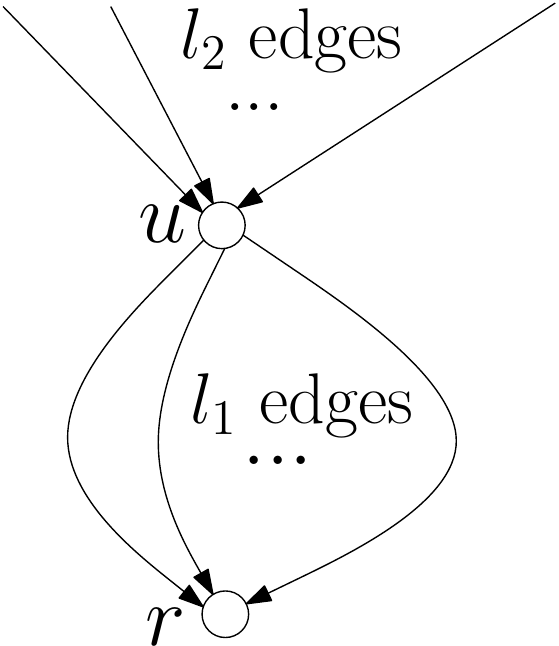}
 \end{center}
\caption{A network reduced from the NCPRLC instance $(H,\SizErr,{\bf
e})$.} \label{CommHard}
\end{figure}

Since $H$ is a matrix chosen uniformly at random over $\Field_q$, it
corresponds to a RLNC, where each column of
$H$ corresponds to an IRV of an edge in $\Incoming(u)$.

Let $e$ be an edge whose tail is connected to $\SizErr$
edges in $\Incoming(u)$. Let the  IRV of $e$ be ${\bf e}$. If the receiver $r$
can recover the topology, $r$ is able to tell how the tail of $e$ is
connected to the $\SizErr$ edges in $\Incoming(u)$.  Thus $r$ can find
a linear combination of $\SizErr$ columns of $H$ resulting in ${\bf
e}$ and thus solve $(H,\SizErr,{\bf e})$. \hfill $\Box$

\subsection{Topology estimation for networks with adversarial
errors} \label{subsec:Topo-adv-RLNC}

 In this section, we use an
error-correcting code approach~\cite{RankMetricRandCode} to estimate
the topology of a network with adversarial errors. At a high
level, the idea is that in strongly connected networks,
each pair of networks generates transform matrices that look ``very
different". Hence no matter what the adversary does, he is unable to
make the transform matrix for one network resemble that of any
other. The estimation algorithm and proof techniques are similar in
flavor to those from algebraic coding theory.

As is common in the network error-correcting literature, we assume
that the adversary is bounded, and therefore corrupts no more than
$\SizErr$ edges in the network.

\noindent {\bf Assumptions and justifications:}
\begin{enumerate}
\item At most $\SizErr$ edges in $\ErrEdg$ suffer errors, {\it i.e.},
$\{e: e\in \Edges, \EdgInj(e)\neq 0\}=\ErrEdg$ and $|\ErrEdg|\leq
\SizErr$. When $2\SizErr+1\leq \Capacity$, network-error-correcting
codes (see Section~\ref{ErrorCorrectCode} for details) are used so
that the source message $\eX$ is provably decodable.

\item {\it Strong connectivity}. A set of networks satisfies ``strong connectivity" if the following is true: each internal node has both in-degree and out-degree at least
$2\SizErr+1$.
Note that in an acyclic graph it implies the source has at least
$2\SizErr+1$ edge disjoint pathes to each internal node, which has
$2\SizErr+1$ edge disjoint pathes to the receiver. We motivate this
strong connectivity requirement by showing in
Theorem~\ref{thm:net_conn_necc} lower bounds on the connectivity
required for {\it any} topology estimation scheme to work in the
presence of an adversary.\footnote{Note that for the single source (or single receiver) network, such connectivity requires parallel
edges at the source (or the receiver). Otherwise if parallel edges are not allowed, we assume the neighbors of the
source (or the receiver) are the end-nodes, {\it i.e.}, they are not in the domain of tomography. Similar argument holds in later sections.}\label{ass:adv_strongconn}

\item \label{ass:local_topo} {\it Knowledge of local topology.} We assume
that each node knows the ID numbers of the nodes exactly one hop
away from it, either upstream or downstream of it.

\item {\it Strong type common randomness is assumed}. It is justified by Theorem~\ref{thm:comm_rand_necc}.

\end{enumerate}

After receiving the overall transform matrix $T_e$ which is polluted
by the network adversarial errors, the receiver $r$ using the
following algorithm to estimate of topology of the network.

\begin{itemize}
\item {\bf ALGORITHM I: TOPO-ADV-RLNC}: Under RLNC, the algorithm is to use the end
information observed by the receiver to estimate the network topology in the presence of network adversaries.

\item The {\it inputs} are $T_e$ and $\mathcal R=\{\mathcal R_v, v\in \Nodes\}$.
The {\it output} is a graph $\Graph$.

\item Step A: For each candidate graph $\Graph$ with nodes in
$\Nodes$ and satisfying the strong connectivity requirement (see
Assumption~\ref{ass:adv_strongconn}) for details), goto Step B.

\item Step B: Using $\mathcal R$, receiver
$r$ computes the overall transform matrix $\Tran(\Graph)$ for
$\Graph$. If rank$(\Tran(\Graph)-\Tran_e)\leq \SizErr$, output
$\Graph$ and goto Step C; otherwise, go back to continue the loop in Step A.

\item Step C: End {\bf TOPO-ADV-RLNC}.
\end{itemize}

Before proving the correctness of {\bf TOPO-ADV-RLNC} we show the key
lemma for the \emph{rank distance} of different graphes. The {\it
rank-distance} between any two matrices $A, B\in
\Field_q^{\Capacity\times \Capacity}$ is defined as
$r_m(A,B)=\mbox{rank}(A-B)$. We note that rank-distance indeed satisfies
the properties of a distance function; in particular it satisfies
the triangle inequality~\cite{RankMetricRandCode}.

\begin{lemma}
\label{LeRankMec} Let the transform matrices of different networks
$\Graph$ and $\Graph'$ be $\Tran(\Graph)$ and $\Tran(\Graph')$
respectively. Then with a probability at least $1-{|\Nodes|^4}/{q}$,
$r_m(\Tran(\Graph),\Tran(\Graph'))\geq 2\SizErr+1$.
\end{lemma}

\noindent{\bf Proof:} Since $\Graph\neq \Graph'$, there exists a
node $u\neq r$ in $\Graph$ which is either not in $\Graph'$ or has an
outgoing edge $e_u$ in $\Graph$ but not in $\Graph'$.\footnote{Otherwise we can
switch the roles of $\Graph$ and $\Graph'$ in the proof.}

We first show that there exists a $(2\SizErr+1)\times (2\SizErr+1)$
sub-matrix in $\Tran(\Graph)-\Tran(\Graph')$, such that its
determinant is not zero on an evaluation of the
elements of the code-books in $\mathcal R$.
Using Assumption~\ref{ass:adv_strongconn}), in $\Graph$ there exist
$2\SizErr+1$ edge disjoint pathes from $s$ to $r$ via $u$. The elements in $\mathcal R$
can be evaluated such that i) only the routing
transmissions along these pathes are allowed; ii) in $\Graph$, the
source $s$ can transmit $2\SizErr+1$ packets using routing via $u$
to $r$; iii) the elements in $\mathcal R_u$ satisfy $\mathcal R(v,u,w,w')=0$ if $(u,w') \neq e_u$.

Thus for graph $\Graph$, under such evaluation of $\mathcal R$ the
transform matrix $\Tran(\Graph)$ has a sub-matrix as a
$(2\SizErr+1)\times (2\SizErr+1)$ identity matrix.

For the case where $u\not \in \Graph'$, since the receiver $r$ can
only receive the routing transmissions via $u$, the transform matrix
$\Tran(\Graph')$ is therefor a zero matrix.

For the case where $u\in \Graph'$, since $e_u\not \in \Graph'$, in $\Graph'$ all local coding coefficients used
by $u$ are zero and therefor the
transform matrix $\Tran(\Graph')$ is still a zero matrix.

Thus under such evaluation of $\mathcal R$,
$\Tran(\Graph)-\Tran(\Graph')$ always has a $(2\SizErr+1)\times
(2\SizErr+1)$ sub-matrix with determinant $1$.

The determinant of the sub-matrix is a polynomial of random variables
belonging to $\mathcal R$, with degree at most $|\Edges|\times
(2\SizErr+1)\leq |\Edges|^2\leq |\Nodes|^4$. Using Schwarth-Zippel
Lemma~\cite{CCBook}, with a probability at least
$1-\frac{|\Nodes|^4}{q}$ the determinant of the sub-matrix is
nonzero, i.e., $r_m(\Tran(\Graph)-\Tran(\Graph'))\geq 2\SizErr+1$.
\hfill $\Box$

Since there are at most $2^{|\Nodes|^2/2}$ acyclic graphs and
$2^{|\Nodes|^2}$ pairs of graphs to be compared, following a Union Bound~\cite{ProbabilityBook2005}
argument, with a
probability at least $1-{|\Nodes|^42^{|\Nodes|^2}}/{q}$ the lemma is true for any pair of networks \footnote{For counting the total number of
networks we do not count the the networks with parallel edges for
clarity of exposition. When parallel edges is taken into count, the
length of field size $q$ should be $\Theta(|\Nodes|^2\log(|\Edges|))$ to make the failure probability of
tomography negligible. }.

As in (\ref{eq:net_tran_err}), after transmission, the erroneous
transfer matrix $\Tran_e$ received by $r$ is actually
\begin{eqnarray}
\Tran_e=\Tran+\IRM (\ErrEdg) \Errors_h, \label{eq:err_head}
\end{eqnarray}
where $\Errors_h$ represents the errors injected for
the packet headers, {\it i.e.}, the first $\Capacity$ columns of
$\Errors$. This combined with Lemma~\ref{LeRankMec} enables us to
prove the correctness of {\bf TOPO-ADV-RLNC}.

\begin{theorem}
\label{thm:adv-topo-recover} With a probability at least
$1-|\Nodes|^42^{|\Nodes|^2}/q$, the network $\Graph$ outputted by
{\bf TOPO-ADV-RLNC} is the correct network. \label{thm:adv_tomo}
\end{theorem}
\noindent {\bf Proof: } We assume lemma $\ref{LeRankMec}$ is true
for any pair of graphs, which happens with a probability at least
$1-|\Nodes|^42^{|\Nodes|^2}/q$ as stated above.

By (\ref{eq:err_head}), the rank distance
$r_m(\Tran_e,\Tran(\Graph))$ equals $rank(\IRM (\ErrEdg)
\Errors_h)\leq rank(\IRM (\ErrEdg))\leq \SizErr$. For any transfer matrix
$\Tran(\Graph')$ corresponding to a different network $\Graph'$, by
the triangle inequality of the rank distance,
$r_m(\Tran(\Graph'),\Tran_e)\geq
r_m(\Tran(\Graph'),\Tran(\Graph))-r_m(\Tran(\Graph),\Tran_e)\geq
\SizErr+1$. This completes the proof. \hfill $\Box$

In the end, we show that the strong connectivity requirements (see
Assumption~\ref{ass:adv_strongconn}) for details) we require for
Theorem~\ref{thm:adv_tomo} are ``almost'' tight\footnote{We remark
that there is a mismatch between the sufficient connectivity
requirement in Assumption~\ref{ass:adv_strongconn}) (that there be
$2\SizErr+1$ edges between $s$ and each node), and the necessary
connectivity requirement of Theorem~\ref{thm:net_conn_necc}
 (that there be $\SizErr+1$ edges between $s$ and each node). Whether the gap between such mismatch can be closed is still open.
}.
\begin{theorem}
\label{thm:adv-topo-conn} For any network $\Graph$ that has fewer
than $\SizErr+1$ edges from the source $s$ to each node, or fewer
than $2\SizErr+1$ edges from each node to the receiver $r$, there
exists an adversarial action that makes any tomographic scheme fail
to estimate the network topology. \label{thm:net_conn_necc}
\end{theorem}
\noindent {\bf Proof:} Assume node $v$ has a min-cut $2\SizErr$ to the
the receiver $r$, and the adversary controls a set $\ErrEdg$ of size
$\SizErr$ of them. When the adversary runs a fake version of the tomographic
protocol announcing that $v$ is not connected to the edges in
$\ErrEdg$, the probability that $r$ incorrectly infers the presence
of $v$ is $1/2$.

On the other hand, if $v$ has only $\SizErr$ incoming edges, the
adversary can cut these off ({\it i.e.} simulate erasures on these
edges). Since the node can only transmit the message from its
incoming edges, this implies that all messages outgoing from $u$ are
also, essentially, erased. Hence the presence of $v$ cannot be
detected by $r$. \hfill $\Box$

In fact, the proof of Lemma~\ref{LeRankMec} only requires $\Graph$
and $\Graph'$ differs at a node
 with high connectivity. If we know the possible
topology set a priori, we can relax the connectivity requirement.
The following corollary formalizes the observation.
\begin{corollary}
\label{Cor:Partial-topo-adv} For a set of possible networks
$\{\Graph_1,\Graph_2,...,\Graph_d\}$, if any two of them differs at
a node which has max-flow at least $2\SizErr+1$ from the source and
max-flow at least $2\SizErr+1$ to the receiver, with a
probability at least $1-d^2|\Nodes|^4/q$ the receiver can
find the correct topology by the receiving transform matrix.
\end{corollary}

\subsection{Topology estimation for networks with random failures}

\label{subsec:PTTopo}  Under RLNC, we provide a polynomial-time scheme to
recover the topology of the network that suffers random network
errors (the definition of random errors can be found in Section~\ref{subsec:errormodel}). The receiver $r$ proceeds in two stages. In the first stage
({\bf Algorithm II: FIND-IRV}), $r$ recovers the IRV information during several
rounds of network communications suffering random errors. In the
second stage ({\bf Algorithm III: FIND-TOPO}), $r$ uses the IRV
information obtained to recover the topology. An interesting feature
of the algorithms proposed is that random network failures actually
make it {\it easier} to efficiently estimate the topology.

\noindent {\bf Assumptions, justifications, and notation:}
\begin{enumerate}
\item { Multiple ``successful'' source generations. A ``successful'' generation  means the number of
errors does not exceed the bound $C-1$ and receiver $r$ can decode the
source message correctly using network error-correcting-codes (see
Section~\ref{ErrorCorrectCode} for details)}. The protocol runs for
$t$ independent ``successful'' source generations, where $t$ is a
design parameter chosen to trade off between the probability of
success and the computational complexity of the topology estimation
protocol. Let $X(i)$ be the source messages transmitted,
$\ErrEdg(i)$ be the edge set suffering errors and $\RevMat(i)$ be
the received matrix in the $i$th source generation.

\item \label{ass:weak_connect} { Weak connectivity requirement.} It is assumed that
each internal node has out-degree no less than $2$.
 Note it is the necessary condition that each edge
is {\it distinguishable} from every other edge, {\it i.e.}, any pair
of edges are flow-independent (see the definition in
Section~\ref{subsec:dependency} for details).

\item { Each node knows the IDs of its neighbors.} As in Section~\ref{subsec:Topo-adv-RLNC}, Assumption~\ref{ass:local_topo}).

\item \label{ass:ind_fail} The network is not ``noodle like'' ({\it i.e.},
high-depth and narrow-width)\footnote{ At a high-level, the problem
lies in the fact that such networks have high description complexity
(dominated by the height), but can only support a low information
rate (dominated by the width).}. To be precise for any distinct
$i,j\in [1,t]$ let the random variable $\mathcal D(i,j)$ be $1$ if
and only if $\ErrEdg(i)$ is flow-independent to $\ErrEdg(j)$,
{\it i.e.}, $\mbox{\it flow-rank}(\ErrEdg(i)\cup
\ErrEdg(j))=\mbox{\it flow-rank}(\ErrEdg(i))+\mbox{\it
flow-rank}(\ErrEdg(j))$. Since the random network errors are independent of the source generation, $Pr(\mathcal
D(i,j)\neq 1)$ has no dependence on $(i,j)$ and is defined as $\ProEdgeCorr$. The network is not ``noodle like'' requires $\ProEdgeCorr$ bounded away from $1$.

\item \label{ass:all-fail}For each source generation, each edge $e$
independently has random errors with probability at least $\ProEdgFail$. Note that Assumption 1) and 4) require
 the typical number of error edges $\ProEdgFail|\Edges|$ in each source generation is no more than $C$. Thus we can assume
 $\ProEdgFail=\Theta(1/|\Edges|)$.

\item\label{ass:Weak-Commonrand} {Weak type common randomness is assumed.} It is justified
by Theorem~\ref{thm:high_comp_rand}.
\end{enumerate}

\noindent{\bf Stage I: Find candidate IRVs}

Recall the source message is formed as $X(i)=[I_\Capacity,M(i)]$,
where $I_\Capacity$ is a $\Capacity\times \Capacity$ identity matrix
and $M(i)\in \Field_q^{\Capacity\times(\bl-\Capacity)}$ is the
message. For any matrix $N$ with $\bl$ columns, let $N_h$ (and $N_m$
) be the matrix comprised of the first $C$ columns (and last $n-C$ )
of $N$. Then the algorithm that finds a set of candidate IRVs is as
follows:
\begin{itemize}
\item {\bf Algorithm II, FIND-IRV}: The algorithm is to recover a
set of candidate IRVs of the network from $t$ ``successful'' source
generations.

\item The {\it input} is $\{\RevMat(i),i\in[1,t]\}$. The {\it output} is
$\IRVCand$ which is a set of dimension-one subspaces in
$\Field_q^{C}$ and initialized as an empty set.

\item Step A: For $i\in [1,t]$, $r$ computes $M(i)$ using network error-correction-code (see
Section~\ref{ErrorCorrectCode} for details) and then
$\ErrorsMat(i)_r=\RevMat(i)_m-\RevMat(i)_h M(i)$.

\item Step B: The intersection of the column-spaces ${\bf
\ErrorsMat(i)_r}\cap{\bf \ErrorsMat(j)_r}$ is computed for each pair
$i, j \in \{1,\ldots,{t}\}$. If $rank({\bf \ErrorsMat(i)_r}\cap{\bf
\ErrorsMat(j)_r})=1$ for any $(i,j)$ pair, ${\bf
\ErrorsMat(i)_r}\cap{\bf \ErrorsMat(j)_r}$ is added into $\IRVCand$.

\item Step C: End {\bf FIND-IRV}:
\end{itemize}

Let $\ProNonAcp$ denote
$\ProEdgeCorr+2\ProRandomNonSpan+{|\Edges|}/{q}$ and
$\ProRandomNonSpan$ be
$1-(1-{\SizErr}/{q})[1-2{\Capacity^2}/(\bl-\Capacity)]$ and $<{\bf v}>$ be the dimension-one subspace spanned by any vector ${\bf v}$. Then the
theorem followed characterizes the performance of {\bf FIND-IRV}.

\begin{theorem}
The probability that $\IRVCand$ contains $\{<\IRV {e}>: e\in \Edges\}$ is at least
$1-|\Edges| \ProNonAcp^{{{t} \ProEdgFail}/{2}}$.
\label{LeNodeTranAcc}
\end{theorem}

The proof will be presented later.

\noindent{\bf Remark $1$}: Each element in $\IRVCand$ has no correspondence with any edge in the network. Such
correspondences would be found in next stage by algorithm {\bf FIND-TOPO}.

\noindent{\bf Remark $2$}: The probability $\ProRandomNonSpan$ asymptotically
approaches $0$ with increasing block-length-$\bl$ and field-size-$q$.
Hence $\ProNonAcp$ is bounded away from $1$ using
Assumption~\ref{ass:ind_fail}). Thus if $t= \Theta (\log(|\Edges|)/\ProEdgFail)$, the probability that
$\IRVCand$ contains $\{<\IRV {e}>: e\in \Edges\}$ is $1-o(1)$. Since $\ProEdgFail=\Theta(1/|\Edges|)$, without
loss of generality we henceforth assume $t=\Theta(|\Edges|\log(|\Edges|))$.

\noindent{\bf Remark $3$}: Since $2\ProRandomNonSpan+{|\Edges|}/{q}$ is
asymptotically negligible for large block-length $\bl$ and field
size $q$, $\ProNonAcp$ approximately equals $\ProEdgeCorr$. Also
Lemma~\ref{LeIRVandEdges} and Lemma~\ref{LeRandomErr} imply that for
large $\bl$ and $q$, any two failing edge-sets $\ErrEdg(i)$ and
$\ErrEdg(j)$ across multiple source generations are
flow-independent if and only if the corresponding error matrices ${
\ErrorsMat(i)_r}$ and ${ \ErrorsMat(j)_r}$ are column linearly
independent. Thus $r$ can estimate $1-\ProNonAcp$ and hence
$1-\ProEdgeCorr$ by estimating the probability that pairs of ${\bf
\ErrorsMat(i)_r}$ and ${\bf \ErrorsMat(j)_r}$ are linearly
independent. This enables $r$ to decide how many communication
rounds $t$ are needed so that {\bf FIND-IRV} has the desired probability
of success.

\begin{figure}
\begin{center}
 \includegraphics[width=30mm]{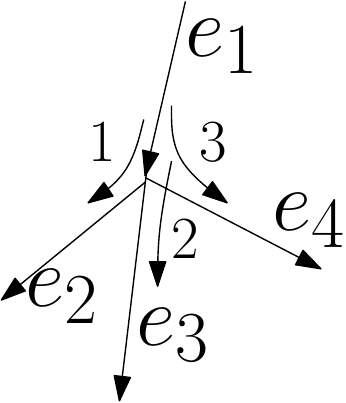}
 \end{center}
\caption{\label{FirstExample}Let $\ErrEdg(1)=\{e_1,e_4\}$ and
$\ErrEdg(2)=\{e_2,e_3\}$ and
$rank(\IRV{e_2},\IRV{e_3},\IRV{e_4})=3$, then we have $rank({\bf
\IRM(e_2,e_3)}\cap{\bf \IRM(e_1,e_4)})=1$ and ${\bf
\IRM(e_2,e_3)}\cap{\bf \IRM(e_1,e_4)}={\bf [\IRV{e_2}+2\IRV
{e_3}]}$, which is a ``fake candidate''.} \label{FakeCand}
\end{figure}

\noindent{\bf Remark $4$}: The set of vectors output by {\bf FIND-IRV} can also
include some ``fake candidate", as demonstrated in the example in Figure
\ref{FakeCand}. In the next stage for topology estimation, these
fake IRVs will be filtered out automatically.

Before the proof of Theorem~\ref{LeNodeTranAcc}, we show the
following lemma arisen from the properties of random errors and is a
core lemma for network tomography in random network errors.

\begin{lemma}
\label{LeRandomErr} For random error model, ${\bf
\ErrorsMat(i)_r}={\bf \IRM(\ErrEdg(i))}$ with a probability at least
$1-\ProRandomNonSpan$.
\end{lemma}
\noindent{\bf Proof:} Recall that $\Errors(i)_m$ comprised of the
last $\bl-\Capacity$ columns of $\Errors$. We first prove that
$\Errors(i)_m$ has full row rank $\SizErr$ with a probability at least
$1-\ProRandomNonSpan$.

In the random error model (see Section~\ref{subsec:errormodel} for
details) each error edge $e$ has at least one randomly chosen location (say $\ell$)
in the injected packet ${\bf z}(e)$ such that the $\ell$th component of ${\bf z}(e)$ is  chosen uniformly at random  from
$\Field_q$. Thus for each
row of $\Errors(i)$, all the last $\bl-\Capacity$ elements are zero
with a probability at most $\Capacity/\bl$. Using Union Bound~\cite{ProbabilityBook2005}
$\Errors(i)_m$ has zero rows with a probability at most
$\Capacity^2/\bl$. Thus in the following we assume each row of $\Errors(i)_m$ is
non-zero.

 The ``Birthday
Paradox''~\cite{ProbabilityBook2005} implies that with a probability at least
$1-{\Capacity^2}/(\bl-\Capacity)$, for each row of $\Errors(i)_m$,
the following happens: there are $\SizErr$ {\it distinct} column
indexes $l_1,\ldots,l_\SizErr \in\{1,\ldots,\bl-\Capacity\}$ such
that $\Errors(i)_m(i,l_i)$
is chosen uniformly at random. Then the determinant of the
sub-matrix of the $\{l_1,\ldots,l_\SizErr\}$th columns of
$\Errors(i)_m$ is a nonzero polynomial of degree $\SizErr$ of
uniformly random variables over $\Field_q$. By the Schwartz-Zippel
Lemma~\cite{CCBook} this determinant is non-zero with a probability at
least $(1-{\SizErr}/{q})$. Thus $\Errors(i)_m$ has $\SizErr$
independent columns with a probability at least
$(1-{\SizErr}/{q})[1-{2\Capacity^2}/({\bl-\Capacity})] =
1-\ProRandomNonSpan$.

Since $\ErrorsMat(i)_r=\ErrorsMat(i)_m-\ErrorsMat(i)_h
M(i)=\IRM(\ErrEdg(i)) (\Errors(i)_m-\Errors(i)_h M(i))$ and the
non-zero random variables in $\Errors(i)_m$ are chosen independently
from $\Errors(i)_h M(i)$, $(\Errors(i)_m-\Errors(i)_h M(i))$ has
full row rank $\SizErr$ with the same probability
$1-\ProRandomNonSpan$. Thus ${\bf \ErrorsMat(i)_r}={\bf
\IRM(\ErrEdg(i))}$ with a probability at least $1-\ProRandomNonSpan$.
 \hfill $\Box$

Then we have:

 \noindent{\bf Proof of Theorem~\ref{LeNodeTranAcc}:} For any edge $e$ and any $i, j
\in \{1,\ldots,t\}$ and $e\in \ErrEdg(i)\cap \ErrEdg(j)$, we compute the probability
of the event ${\cal F}(e,i,j)$: $<{\bf \IRV {e}}>$ equals the one-dimensional subspace ${\bf \ErrorsMat(i)_r}\cap{\bf \ErrorsMat(j)_r}$.

By Assumption~\ref{ass:ind_fail}), with a probability at least
$1-\ProEdgeCorr$, $\ErrEdg(i)-e$ is flow-independent of
$\ErrEdg(j)-e$. Conditioned on this,
Lemma~\ref{LeIRVandEdges}.\ref{LeIndIRVEdg} implies that with a probability at least $1-\ProEdgeCorr-{|\Edges|}/{q}$, ${\bf
\IRM(\ErrEdg(i) \backslash e)}$ is linearly independent of ${\bf
\IRM(\ErrEdg(j) \backslash e)}$. Hence ${\bf
\IRM(\ErrEdg(i))}\cap{\bf \IRM(\ErrEdg(j))}$ equals
$<{\bf \IRV {e}}>$. And by Lemma~\ref{LeRandomErr}, either of  ${\bf
\ErrorsMat(i)_r}\neq {\bf \IRM(\ErrEdg(i))}$ or  ${\bf
\ErrorsMat(j)_r}\neq {\bf \IRM(\ErrEdg(j))}$ with a probability at
most $\ProRandomNonSpan$. Conditioning on all the events implies
that the probability of event ${\cal F}(e,i,j)$ is at least
$1-\ProEdgeCorr-2\ProRandomNonSpan-{|\Edges|}/{q}$.

When ${t}$ is large enough, by the Chernoff bound~\cite{ProbabilityBook2005} $e$
will fail at least $t\ProEdgFail/2$ times with a probability at least
$1-\ProEdgFail^{\Theta(t) }$. Conditioned on these many failures,
there are ${t \ProEdgFail}/4$ probabilistically independent ${\cal
F}(e,i,j)$ for edge $e$, and {\bf FIND-IRV} accepts $\IRV {e}$ with a probability at
least $1-(\ProNonAcp^{{{t} \ProEdgFail}/{4}}+\ProEdgFail^{\Theta(t) }$). Taking the Union Bound over all edges gives the required
result.
\hfill $\Box$

\noindent{\bf Stage II: Topology recovery via candidate IRVs}

Using $\IRVCand$, we now describe {\bf Algorithm FIND-TOPO} that
determines the network topology.

Note that $\IRVCand$ is merely a set of dimension-one subspaces,
and as such, individual element may have no correspondence with the
actual IRV of any edge in the network. At any point in {\bf
FIND-TOPO}, let $\FindGraph$ denote the network topology recovered
thus far. Let $\FindNode$ and $\FindEdge$ be the corresponding sets
of nodes and edges respectively in $\FindGraph$, and $\FindIRV$ be
the set of IRVs of the edges in $\FindEdge$, which are computed from
$\FindGraph$ and the set of local random code-books $\mathcal R=\{\mathcal R_v:v\in \Nodes\}$. We note that the
IRVs in $\FindIRV$ are vectors rather than dimension-one
subspaces.

We describe algorithm estimating the network topology as follows.
\begin{itemize}
\item {\bf Algorithm III, FIND-TOPO}: The algorithm is to use
$\IRVCand$ and $\mathcal R=\{\mathcal R_v, v\in \Nodes\}$ to recover
the network topology.

\item The {\it input} is $\IRVCand$ and
$\mathcal R$. The {\it output} is $\FindGraph =
(\FindNode,\FindEdge)$.

\item  Step A: The set $\FindNode$ is initialized as the receiver $r$, all its  upstream
neighbors, and the source $s$. The set $\FindEdge$ is initialized as
the set of edges incoming to $r$. Hence $\FindGraph =
(\FindNode,\FindEdge)$. The initial set of $\FindIRV$ are the IRVs
of the incoming edges of $r$, {\it i.e.}, a set of distinct columns of the $C\times C$ identity matrix. The state flag STATE(New-Edge) is initialized to be ``False''.

\item Step B: For each node $v\neq s$ in $\FindNode$, call function $FindNewEdge(v)$ (Step C).
If
\begin{itemize}
\item STATE(New-Edge) is ``True'', set STATE(New-Edge) be ``False'' and repeat the loop of Step B.

\item STATE(New-Edge) is ``False'', go to Step E.
\end{itemize}

\item Step C: (Function $FindEdge(v)$) Let $e_1,\ldots,e_d$ be the
outgoing edges of $v$ in $\FindGraph$. If
$\{\FindIrv{e_1},\ldots,\FindIrv{e_d}\}$ from $\FindIRV$ has
\begin{itemize}
\item rank $1$, step-back and continue the loop in Step B.

\item rank greater than $1$, for
each candidate incoming edge of $v$, say $e=(u,v)$, if
$e\not\in\FindEdge$, call function $CheckIRV(v,e)$ (Step D).
Step-back and continue the loop in Step B.
\end{itemize}

\item Step D:  (Function $CheckIRV(v,e)$)
Use ${\cal R}$ to compute the IRV of $e$ as
$\FindIrv{e}=\sum^d_{j=1}\beta(e,v,e_j) \FindIrv{e_j}$. Check
whether $<{\FindIrv{e}}>$ is in $\IRVCand$. If so,
\begin{enumerate}
\item Set STATE(New-Edge) be ``True''.

\item If $u\not\in\FindNode$, add $u$ to $\FindNode$.

\item Add $e=e(u,v)$ to $\FindEdge$.

\item Based on $\mathcal R$, update $\FindIRV$ from $\FindGraph=(\FindNode,\FindEdge)$.\footnote{
The reason that $\FindIRV$ needs to be updated is that: when $e$ is
found as a new edge in $\FindGraph$, the IRVs of the edges upstream
of $e$ in $\FindGraph$ will change.}.
\end{enumerate}
Step-back to the loop in Step C.

\item Step E: End {\bf FIND-TOPO}.
\end{itemize}

If $\IRVCand$ contains all edge IRVs which is supported by
Theorem~\ref{LeNodeTranAcc}, we show correctness of {\bf FIND-TOPO}
as:
\begin{theorem}
With a probability $1-{\mathcal O(\log^2(|\Edges|)|\Edges|^4|\Nodes|)}/{q}$, {\bf
FIND-TOPO} recovers the accurate topology by performing $\mathcal
O(\log^2(|\Edges|)|\Edges|^4|\Nodes|\Capacity)$ operations over
$\Field_q$.\label{Th:FIND-Topo-Rand}
\end{theorem}

Before the proof of Theorem~\ref{Th:FIND-Topo-Rand} we need the
following lemma, which shows that with high probability function
$CheckIRV(v,e)$ accept an edge $e$ if and only if $e$ is actually in
the network $\Graph$.

\begin{lemma}
\label{LeCorTopo} \begin{enumerate}
\item If edge $e = (u,v)$ exists in $\Graph$,
$<\IRV{e}>$ is in $\IRVCand$, $\{e_1,\ldots,e_d\}$ are exactly all
the outgoing edges of $v$ in $\Graph$ and $\FindIrv{e_i}=\IRV{e_i}$
for $i=1,2,...,d$, function $CheckIRV(v,e)$ accepts $e$ as a new
edge in $\FindEdge$ with a probability $1$.

\item If edge $e$ does not exist
in $\Graph$, function $CheckIRV(v,e)$ accepts $e$ as a new edge in
$\FindEdge$ with a probability ${\mathcal O(\log^2(|\Edges|)|\Edges|^2)}/{q}$.
\end{enumerate}
\end{lemma}
\noindent{\bf Proof:}
\begin{enumerate}
\item Under the conditions we have
$\FindIrv{e}=\sum^d_{j=1}\beta(e,v,e_j) \FindIrv{e_j}=\IRV{e}$ and
will be accepted.

\item If $e$ does not exist in $\Graph$, the coding coefficients $\{\beta(e,v,e_j) : j=1,\ldots,d\}$ are not used. Hence
from the perspective of any element $<{\bf h}>$ in $\IRVCand$,
$\sum^d_{j=1}\beta(e,v,e_j)  \FindIrv{e_j}$ is an independently and
uniformly chosen vector in the span of the vectors ${\bf
\{\FindIrv{e_j}:j \in \{1,\ldots,d\}\}}$. Since $CheckIRV(v,e)$ is
called only if the rank of $\{{\bf \FindIrv{e_j}:j \in
\{1,\ldots,d\}}\}$ is no less than $2$,  so that ${\FindIrv {e}}\in <{\bf
h}>$  with a probability at most ${1}/{q}$.
Since {\bf FIND-IRV} in Stage I needs at most $t = \mathcal
O(\log(|\Edges|)|\Edges|)$ source generations\footnote{As pointed out in Remark $2$ after
Theorem~\ref{LeNodeTranAcc}.}, $\IRVCand$ has size at most $\mathcal
O(\log^2(|\Edges|)|\Edges|^2)$. Using the Union Bound~\cite{ProbabilityBook2005} $<{\FindIrv {e(u,v,i)}}>$ is in $\IRVCand$ with a probability $\mathcal O(\log^2(|\Edges|)|\Edges|^2/q)$. \hfill $\Box$
\end{enumerate}

Then we have:

 \noindent{\bf Proof of Theorem~\ref{Th:FIND-Topo-Rand}:} Note that if
no errors occur, Step B can find at most $|\Edges|$ edges, each time
of finding a new edge of Step B needs at most $|\Nodes|$ invocations
of Step C (once for each node), and each invocation of Step C
results in at most $|\Edges|$ invocations of Step D. Thus Step D can
be invoked at most $|\Edges|^2|\Nodes|$ times, and
Lemma~\ref{LeCorTopo} demonstrates that each invocation results in
an error with a probability at most ${\mathcal O}{(\log^2(|\Edges|)|\Edges|^2/q)}$.
Note further that this is the only possible error event. Hence by
the Union Bound~\cite{ProbabilityBook2005} the probability that {\bf FIND-TOPO}
results in an erroneous reconstruction of $\Graph$ is ${\mathcal
O(\log^2(|\Edges|)|\Edges|^4|\Nodes|)}/{q}$. Also, each computation of Step D takes
at most ${\cal O}(\log^2(|\Edges|)|\Edges|^2\Capacity)$ finite field comparisons to
determine membership of $<\FindIrv{e}>$ in
$\IRVCand$. Hence, given that the bound on the number of invocations
of Step D and that this can be verified to be the most
computationally expensive step, the running-time of {\bf FIND-TOPO}
is $O(\log^2(|\Edges|)|\Edges|^4|\Nodes|\Capacity)$ operations over $\Field_q$.

Finally, we note that $\Graph$ is acyclic and the assumption that
$\IRVCand$ contains $\{<\IRV {e}>: e\in \Edges\}$. Hence conditioning on no
incorrect edges being accepted, for each invocation of Step B,
unless $\FindGraph=\Graph$, there exists an edge $e$ such that all
edges $e'$ downstream of $e$ in $\Graph$ are in $\FindEdge$, which
implies all the corresponding $\FindIrv{e'}$s are correctly
computed. Thus by Lemma~\ref{LeCorTopo} edge $e$ is accepted into
$\FindEdge$
 by function $CheckIRV(v,e)$ with a probability $1$. Hence,  each edge
 actually in $\Graph$ also eventually ends up in $\FindGraph$, and {\bf FIND-TOPO} terminates.
\hfill $\Box$

%% file: LocateErrforRLNC.tex
\section{Error localization for RLNC}
\label{sec:loc-err-RLNC} As previous works
(~\cite{Tracyloss1,FragouliTopoAllerton2006,FrauglilossrateGlobecom2007}, under RLNC the receiver $r$ must know the network
topology and local random coding coefficients to locate network errors.
Thus in this section receiver $r$ is assumed to know the IRVs of each
edge, which can follow topology estimation algorithms in Section~\ref{sec:Topo-RLNC}, or
the network design as {\it a priori}.

\subsection{Locating adversarial errors under RLNC} \label{sec:locate-adv-error}

In this subsection we demonstrate how to detect the edges in the network where the adversary injects errors. 
Since the IRV is the fingerprint of the corresponding edge,
detecting the error edges thus becomes an equivalent mathematical
problem which detects the IRVs in the error matrix $\ErrorsMat$. Our
technique is based on the fact that when the edges are
flow-independent (see the definition in Section~\ref{subsec:dependency}
for details) enough to each other, the IRV of each error edge is not
erasable from the column space of the error matrix $\ErrorsMat$, {\it i.e.}, ${\bf \ErrorsMat}$.

\noindent{\bf Assumptions  and justifications:}
\begin{enumerate}
\item \label{ass:2} { Each internal node has out-degree at least $2\SizErr$}. Since $\Graph$ is acyclic, it implies that every set of $2\SizErr$
edges in $\Graph$ are flow-independent. While this assumption seems
strong, we demonstrate in Theorem~\ref{thm:conn_adv} that such a
condition is necessary for $r$ to identify the locations of $\SizErr$
corrupted edges.

\item At most $\SizErr$ edges in $\ErrEdg$ suffer errors, {\it i.e.},
$\{e: e\in \Edges, \EdgInj(e)\neq 0\}=\ErrEdg$ and $|\ErrEdg|\leq
\SizErr$. When $2\SizErr+1\leq \Capacity$, network-error-correcting
codes (see Section~\ref{ErrorCorrectCode} for details) are used so
that the source message $\eX$ (and thus the error matrix $\ErrorsMat$) is provably decodable.
\end{enumerate}

Then we have:
\begin{itemize}
\item {\bf ALGORITHM IV, LOCATE-ADVERSARY-RLNC}: The algorithm is to locate the
network adversaries under RLNC.

\item The {\it input} is the error matrix $\ErrorsMat$ and $\{\IRV{e}:e\in \Edges\}$. The {\it output} is a set of edges
$\ErrEdg'$.

\item Step A: Compute rank(${\ErrorsMat}$)$=\RkErr$. Let $\{{\mathbf{e_1}},{\bf
e_2},...,{\bf e_\RkErr}\}$ be a set of independent columns of
${\ErrorsMat}$.

\item Step B: For $i=1,2,...,\RkErr$, find a set of edges
$\ErrEdg_{i}$ with minimal cardinality such that ${\bf e_i}$ is in
the column space of the corresponding impulse response matrix
$\IRM(\ErrEdg_{i})$.

\item Step C: Output $\ErrEdg'=\cup_{i\in[1,\RkErr]}\ErrEdg(i)$.

\item Step D: End {\bf LOCATE-ADVERSARY-RLNC}.
\end{itemize}

We show that with high probability {\bf LOCATE-ADVERSARY-RLNC} finds the
location of edges with adversarial errors.
\begin{theorem}\label{ThFByzanAdverErrors}
With a probability at least $1-{|\Edges|{|\Edges|\choose
{2\SizErr}}}/{q}$ the solution of {\bf LOCATE-ADVERSARY-RLNC} results in
$\ErrEdg'=\ErrEdg$.
\end{theorem}
{\bf Proof:}
 Note that
Assumption 1), with high probability, gives a similar statement about
the rank of the corresponding IRVs. Using the Union Bound~\cite{ProbabilityBook2005} on the result of
Lemma~\ref{LeIRVandEdges}.\ref{LeIndIRVEdg} gives us the
result that any $2\ErrEdg$ IRVs are independent with a probability at
least $1-{|\Edges|{|\Edges|\choose {2\SizErr}}}/{q}$. We henceforth assume it happens in the following.

First of all, since each ${\bf e_i}$ is in ${\bf \IRM(\ErrEdg)}$, we
have $|\ErrEdg_i|\leq \SizErr$ for each $i=1,2,...,\RkErr$.

We claim that for  each $i \in \{1,2,\ldots, \eta\}$, $\ErrEdg_i$
must be a subset of $\ErrEdg$. If not, say $e \in \ErrEdg(i)$ is not
in $\ErrEdg$. By the definition of {\bf LOCATE-ADVERSARY-RLNC}, $\IRV e$ is in the span of the columns of $\IRM(\ErrEdg)$ and
$\IRM(\ErrEdg_i-e)$. Thus a non-trivial combination of the at most
$2\SizErr-1$ IRVs result in $\IRV e$. It contradicts that any
$2\SizErr $ IRVs are linearly independent.

We prove next that for any edge $e\in \ErrEdg$ on which the
adversary injects a non-zero error, {\bf LOCATE-ADVERSARY-RLNC} outputs
at least one $\ErrEdg_i$ such that $e\in \ErrEdg_i$. Without loss of
generality, let $e$ be the first edge in $\ErrEdg$. Then
$\ErrorsMat=\IRM(\ErrEdg)\Errors$ and the first row of $\Errors$ is
nonzero. Since any $\SizErr$ IRVs are independent, $\IRM(\ErrEdg)$
is of full column rank. Then for any $\RkErr$ independent columns in
${\ErrorsMat}$ there must be at least one, say ${\bf e_i}$, such
that the IRV $\IRV{e}$ has nonzero contribution to it. That
is, ${\bf e_i} = \IRM(\ErrEdg)(c_1,c_2,...,c_\SizErr)^T$ with
$c_1\neq 0$. Hence running {\bf LOCATE-ADVERSARY-RLNC} on ${\bf e_i}$
will find $\IRV{e}$ and include the corresponding edge $e$ into
$\ErrEdg_i$. Otherwise, $\IRV {e}$ is in the space of ${\bf \IRM
(\ErrEdg-e, \ErrEdg_{i})}$, which contradicts that any $2\SizErr $
IRVs are linearly independent. \hfill $\Box$

We now show matching converses for
Theorem~\ref{ThFByzanAdverErrors}. In particular, we demonstrate in
Theorem~\ref{thm:conn_adv} that Assumption~\ref{ass:2}), {\it i.e.},
that any $2\SizErr$ edges are flow-independent, is necessary.
\begin{theorem}
\label{thm:high_conn_locate_adv} For linear network coding, any
$\SizErr$ corrupted edges are detectable if and only if any
$2\SizErr$ edges are flow-independent. \label{thm:conn_adv}
\end{theorem}
{\bf Proof:} The ``if'' direction is a corollary of
Theorem~\ref{ThFByzanAdverErrors}. For the ``only if'' direction,
suppose there exist $2\SizErr$ edges such that they are not
flow-independent. Then the corresponding IRVs cannot be linearly
independent by Lemma~\ref{LeIRVandEdges}.\ref{LeCorIRVEdg}. Then
there must exist a partition of these $2\SizErr$ edges into two
edge sets $\ErrEdg_1$ and $\ErrEdg_2$  such
that $|\ErrEdg_1|=\SizErr$ and $|\ErrEdg_2|=\SizErr$ and ${\bf \IRM(\ErrEdg_1)}\cap {\bf \IRM(\ErrEdg_2)}\neq \{0\}$,
{\it i.e.,} the {\it spanning spaces} of the corresponding IRVs in
the two sets intersect non-trivially. Then the adversary can choose
to corrupt $\ErrEdg_1$ and inject errors $\Errors$ in a manner such
that the columns of ${ \IRM(\ErrEdg_1)\Errors}$ are in ${\bf
\IRM(\ErrEdg_2)}$. This means $r$ cannot distinguish whether the
errors are from $\ErrEdg_1$ or $\ErrEdg_2$. \hfill $\Box$

Theorem~\ref{thm:conn_adv} deals with the case that any $\SizErr$
edges can be corrupted. If only some sets of edges are candidates
for adversarial action (for instance the set of outgoing edges from
some ``vulnerable" nodes) we obtain the following corollary.
\begin{corollary}
Let ${\cal S} =  \{\ErrEdg_1,\ErrEdg_2,...,\ErrEdg_t\}$ be
disjoint sets of edges such that exactly one of them is controlled
by an adversary. Then $r$ can detect which edge set is controlled by
the adversary if and only if any two sets $\ErrEdg_i$ and
$\ErrEdg_j$ in ${\cal S}$ are flow-independent.
\end{corollary}

{\it Note:} The flow-independence between edge-sets $\ErrEdg_i$ and
$\ErrEdg_j$ in ${\cal S}$ does not require the edges within
either of $\ErrEdg_i$ or $\ErrEdg_j$ to be flow-independent. It merely
requires that $\mbox{flow-rank}(\ErrEdg_i) +
\mbox{flow-rank}(\ErrEdg_j) = \mbox{flow-rank}(\ErrEdg_i \cup
\ErrEdg_j)$.

Note that running {\bf LOCATE-ADVERSARY-RLNC} might require checking all
the $\Edges\choose \SizErr$ subsets of edges in the network --
this is exponential in $\SizErr$. We now demonstrate that for
networks performing RLNC, the task of
locating the set of adversarial edges is in fact computationally
intractable {\it even when the receiver knows the topology and local
encoding coefficients in advance}.

\begin{theorem}
For RLNC, if knowing the network $\Graph$ and all local coding
coefficients allows the receiver $r$ correctly locating all adversarial
locations in time polynomial in network parameters, NCPRLC (see the
definition in Section~\ref{subsec:NCPHard} for details) can be
solved in time polynomial in problem parameters.
\label{thm:high_comp}
\end{theorem}
{\bf Proof:} Given a NCPRLC instance $(H,\SizErr,{\bf e})$, as shown
in Figure ~\ref{CommHard}, we construct a network with $l_1$ edges
to receiver $r$ and $l_2$ edges to node $u$.

Since $H$ is a matrix chosen uniformly at random over $\Field_q$, it
corresponds to a RLNC, where each column of
$H$ corresponds to an IRV of an incoming edge of $u$.

Assume the adversary corrupts  $\SizErr$ incoming edges
of $u$. Adversary can choose the errors $\Errors$ such that each
column of $\ErrorsMat=\IRM (\ErrEdg)\Errors$ equals ${\bf e}$. In
the mean time $\ErrorsMat$ is all the information about the
adversarial behavior known by $r$ under RLNC. Any algorithm that outputs the
corrupted set $\ErrEdg$ must satisfy ${\bf e}\in {\bf
\IRM(\ErrEdg)}$ and $|\ErrEdg|\leq \SizErr$. Once  $\ErrEdg$ is
found, $r$ actually solves
the NCPRLC instance $(H,\SizErr,{\bf e})$. \hfill $\Box$

\subsection{Locating random errors under RLNC} \label{sec:loc-rad-err} We now
consider the problem of finding the set of edges $\ErrEdg$ that
experience random errors (see Section~\ref{subsec:errormodel} for details). Since ${\bf \IRM}(\ErrEdg)={\bf
\IRM}(\ExtEdgesSetOut(\ErrEdg))$ (see the definition of
$\ExtEdgesSetOut(\ErrEdg)$
 in Section~\ref{subsec:LinTran}
 for reference), the receiver
can not distinguish whether the errors are from $\ErrEdg$ or
$\ExtEdgesSetOut(\ErrEdg)$. So rather than finding $\ErrEdg$, we
provide a computationally tractable algorithm to locate
$\ExtEdgesSetOut(\ErrEdg)$, a proxy for $\ErrEdg$. The algorithm
that finds $\ExtEdgesSetOut(\ErrEdg)$ is as follows:

\begin{itemize}
\item {\bf Algorithm V, LOCATE-RANDOM-RLNC}: Under RLNC, the algorithm is to locate the
edges in the network suffering random errors.

\item The input is the matrix $\RevMat$ received by $r$ and $\{\IRV{e}:e\in \Edges\}$. The output is an edge set $\ErrEdg'$ initialized as an
empty set.

\item Step A: Compute $\ErrorsMat_r$ as the Step A of {\bf Algorithm II: FIND-IRV}.

\item Step B: Check for each edge $e$ whether its IRV $\IRV e$ lies in ${\bf \ErrorsMat_r}$. If so, the edge
$e$ is added into $\ErrEdg'$.

\item Step C: End {\bf LOCATE-RANDOM-RLNC}.
\end{itemize}

The correctness of {\bf LOCATE-RANDOM-RLNC} is followed.
\begin{theorem}
\label{ThFindRandomError} If $\SizErr$ is no more than
$\Capacity-1$, $\ErrEdg'=\ExtEdgesSetOut(\ErrEdg)$ with a probability
at least $1-{3|\Edges|^2}/{q}-2\Capacity^2/(\bl-\Capacity)$. The
computational complexity  is $\mathcal O(|\Edges|\Capacity^2)$
operations over $\Field_q$.
\end{theorem}

\noindent{\bf Proof:} Lemma~\ref{LeIRVandEdges}.\ref{LeIndIRVEdg}
and Lemma~\ref{LeRandomErr} implies that ${\bf \ErrorsMat_r}= {\bf \IRM} (\ErrEdg))={\bf
\IRM}(\ExtEdgesSetOut(\ErrEdg))$ with a probability
at least $1-{2|\Edges|}/{q}-2\Capacity^2/(\bl-\Capacity)$.  It implies $\ExtEdgesSetOut(\ErrEdg)\subseteq \ErrEdg'$.

For the other direction, using the Union Bound~\cite{ProbabilityBook2005} over all $|\Edges|$
edges on Lemma~\ref{LeIRVandEdges}.\ref{LeIndIRVEdg}, with a probability at least $1-{|\Edges|^2}/{q}$, for any edge $e\not\in
\ExtEdgesSetOut(\ErrEdg)$, ${\IRV e}$ is not in ${\bf
\ErrorsMat_r}$. In the end we have $\ExtEdgesSetOut(\ErrEdg)=
\ErrEdg'$ with a probability
at least $1-{3|\Edges|^2}/{q}-2\Capacity^2/(\bl-\Capacity)$.

For each IRV $\IRV e$, it cost at most $\Capacity^2$ operations over
$\Field_q$ to check whether it is in ${\bf \ErrorsMat_r}$. Then the
total complexity of {\bf LOCATE-RANDOM-RLNC} is $\mathcal
O(|\Edges|\Capacity^2)$ operations over $\Field_q$. \hfill $\Box$

%% file: NetworkRSCode.tex
\begin{center}
 {\large {Part II: Design Network Coding for Network Tomography }}\\
\end{center}

\section{Network Reed-Solomon Coding (NRSC)}
\label{Sec:N-RSC}
\subsection{Motivations}
In part I, under random linear network coding (RLNC), network
tomography is studied for both adversarial and random error
models (see Section~\ref{subsec:errormodel} for the
definition of error models). For random error model the schemes for
both {\it static} topology estimation and error localization can be done in
polynomial time, while the schemes for adversarial error model all
cost exponential time. Moreover, under RLNC localizing adversarial
errors is computational intractable (see
Theorem~\ref{thm:high_comp}) and requires the knowledge of network
topology, whose estimation algorithm also costs exponential time.

In this section network Reed-Solomon Coding (NRSC) is proposed to improve the tomographic performance (specially for
 the adversarial error model), and meanwhile
preserving the key advantages of RLNC. To be concrete, NRSC has the
following features:

\begin{itemize}
\item {\it Low implementation complexity}. The proposed NRSC is a linear network coding scheme (see Section~\ref{subsec:lnc} for
details), and can be implemented in a {\it distribute and
efficient} manner where each network node only needs to know the node-IDs of its
adjacent neighbors. Thus once an edge (or node) has left or come, only its adjacent neighbors need to
adjust the coding coefficients.

\item {\it High throughput}. The capacity of multicast is achieved with
high probability.

\item NRSC aids tomography in the following two aspects:
\begin{itemize}
\item {\it Computational efficiency}. For the adversarial error model, the receiver under NRSC can
locate a number of adversarial errors that match a corresponding tomographic upper bound (see Theorem~\ref{thm:high_conn_locate_adv} for
details) in a computationally
efficient manner. For the random error model, a lightweight topology estimation algorithm is provided under NRSC.

\item {\it The robustness for dynamic networks}. For  adversarial (and random) error localization, the algorithms under NRSC do
 not require the priori knowledge of the network topology and thus are robust against edge and node updating. \footnote{Note that under RLNC, the error localization algorithms in previous works~\cite{Tracyloss1,FrauglilossrateAllerton2005,FrauglilossrateGlobecom2007} and this paper require the priori knowledge of the network topology. However, the topology estimation under RLNC costs exponential time for the networks with adversarial errors, and costs polynomial time for the {\it static} networks with random errors.}
 For topology estimation in the random error model, the lightweight algorithm under NRSC fits dynamic networks better than the one under RLNC.
\end{itemize}
\end{itemize}

\subsection{Overview of NRSC}

\begin{figure}
\label{Fig:ToyExampVIRV}
\begin{center}
 \includegraphics[width=45mm,height=30mm]{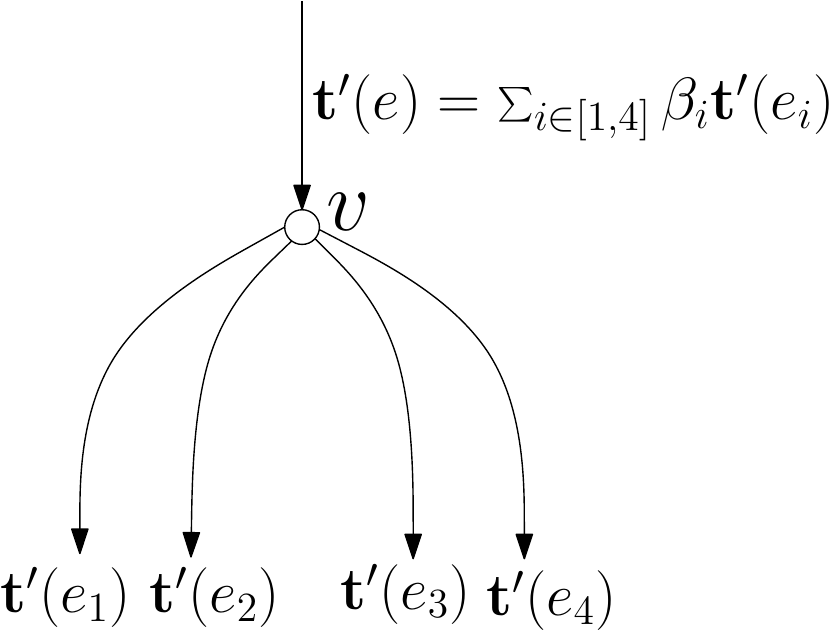}
 \end{center}
\caption{The IRV of $e$ is a linear combination of the IRVs of
$e_1$, $e_2$, $e_3$ and $e_4$.} 
\end{figure}

 In NRSC, in addition
to an IRV each edge $e$ is also assigned a {\it virtual IRV} $\VIRV
{e}$. This virtual IRV is a deterministic function of the node-IDs of
the header and tail of $e$ (and hence is known to them). Further, each
node in an NRSC (say node $v$ in Figure 10) carefully chooses its
coding coefficients ({\it e.g.}, $\{\beta_1 , ..., \beta_4\}$ at
node $v$ in Figure 10, where $\beta_i=\beta(e,v,e_i)$ for $i=1,...,4$) such that the virtual IRVs of edges entering
and leaving $v$ satisfy the same linear relationship as the IRVs
(in the case of Figure~10, $\VIRV {e} = \beta_1
\VIRV{e_1} + ... + \beta_4 \VIRV{e_4}$). In other words, under NRSC every network node makes
``local contribution'' to force edge IRVs equaling the corresponding VIRVs. And the object can be achieved if and only if a connectivity requirement is satisfied (see Corollary~\ref{cor:IRVandVIRV} for details).

At a high level, we compare the tomography performance between RLNC and NRSC in the following:

\begin{itemize}
\item {\it Computational efficiency}. Under RLNC, each edge IRV is randomly chosen from the linear subspace spanned by the down-streaming
edge IRVs, resulting that locating network
adversaries is as hard as NCPRLC (see Theorem~\ref{thm:high_comp} for details). Under NRSC, VIRVs (and then IRVs) are smartly chosen such
that locating adversaries can be done in an efficient manner.

\item {\it The robustness for dynamic networks}. Under RLNC each edge (say $e$) updating results into IRV updating for all up-streaming (of $e$) edges. Under NRSC, once an edge is update, its adjacent header would adjust its local coefficients to stop IRV updating.
For instance consider the subnetwork in Figure~10. Once edge $e_1$ is disconnected,  $v$ would change the local coefficients such that the IRV of $e$ still equals to the VIRV of $e$. Thus no updating is needed for the up-streaming nodes of $v$.
\end{itemize}

\subsection{Node and edge IDs}
\label{subsec:ID-N-RSC} Each pair of nodes $(u,v)$ in $\Nodes\otimes
\Nodes$ has an ID $\ID {u,v}$ chosen independently and uniformly at
random from $\Field_q$. These IDs can be broadcast by the source
using digital signature schemes such as RSA~\cite{Rivest78amethod},
or outputted by a pseudorandom hash function\footnote{Note that the
randomness of the IDs is used in proving Lemma~\ref{Le:EdgeId} and
Theorem~\ref{Th:MulticastCapa}, which (the distinctness of node-pair
IDs and the throughput of multicast) are polynomial time
distinguishable. Thus pseudorandomness
suffices~\cite{ModernCrypto}.} (with input as a pair of nodes) such
as AES~\cite{AESL1} that can be accessed by all parties. Thus this
set of $|\Nodes|^2$ IDs is publicly known {\it a priori} to all
parties (including the adversaries), even though they may not know {\it which} nodes and edges
are actually in the network.

The following lemma shows that each node pair has a distinct ID with
high probability:
\begin{lemma}
\label{Le:EdgeId} With a probability at least $1-|\Nodes|^4/q$, for
any $(u,v)\neq (u',v')$ in $\Edges$, $\ID {u,v}\neq \ID {u',v'}$.
\end{lemma}
\noindent{\bf Proof}: For any $(u,v)\neq (u',v')$, $\ID {u,v}= \ID
{u',v'}$ with a probability at most $1/q$. Since $\Nodes\times \Nodes$
has size $|\Nodes|^2$, there are at most ${|\Nodes|^2 \choose
2}<|\Nodes|^4$ distinct pairs in $\Nodes\times \Nodes$. Using Union Bound~\cite{ProbabilityBook2005} over all these pairs the lemma is
true with a probability at least $1-|\Nodes|^4/q$. \hfill$\Box$

For each edge $e(u,v)\in \Edges$ the ID of $e$ is $\ID {e}=\ID
{u,v}$.  Thus the ID of edge $e(u,v)$ can be figured out by both $u$
and $v$ if they know their adjacent neighbors.
 A direct corollary of Lemma~\ref{Le:EdgeId} is
that each edge has a distinct ID with high probability. We
henceforth assume that this is indeed the case.

Note that for scenario where parallel edges are allowed, we assume
some pairs of nodes has multiple IDs, the $i$'th of which is the ID
of the $i$'th edge between them.

For each edge $e$ the {\it virtual impulse response vector} (VIRV)
is $\VIRV {e}\in \Field_q^C$, which is $[\ID e, (\ID e)^2,...,(\ID
e)^C]^T$. For any set of edges $\ErrEdg$ with size $\SizErr$, the
virtual impulse-response-matrix (VIRM) is $\VIRM (\ErrEdg)\in
\Field_q^{C\times \SizErr}$, with the columns comprised of
$\{\VIRV{e}, e\in \ErrEdg\}$.

For the ease of notation we also defined a dimension-parameterized
 VIRV as $\VIRV {e,i}=[\ID e, (\ID e)^2,...,(\ID e)^i]^T$. For
any set of edges $\ErrEdg$ with size $\SizErr$, the corresponding
VIRM is $\VIRM (\ErrEdg,i)\in \Field_q^{i\times \SizErr}$, with the
columns comprised of $\{\VIRV{e,i}, e\in \ErrEdg\}$. Note that
$\VIRM(\ErrEdg,\SizErr)$ is a Vandermonde matrix and invertible when
$|\ErrEdg|=\SizErr$ and the edges in $\ErrEdg$ have distinct IDs.

\subsection{Code construction of NRSC}
\label{subsec:NRSC-Construct} We assume by default that the edges in
$\Edges$ have distinct IDs, which happens with a probability at least
$1-|\Nodes|^4/q$ by Lemma~\ref{Le:EdgeId}. Recall that $\Capacity$
is the capacity of the network, {\it i.e.},
$\Capacity=\mbox{max-flow} (s,r)$, and for ease of notation we
assume that the source has exactly $\Capacity$ outgoing edges and
the receiver has $\Capacity$ incoming edges (see
Section~\ref{subsec:NetSet} for details).

The construction of NRSC is then as follows.

{\it Source encoder:} Let $\Outgoing
(s)=\{e_1,e_2,...,e_{\Capacity}\}$ be the outgoing edges of the
source $s$ and $\eX\in \Field_q^{\Capacity\times\bl}$ be the source
message matrix. The source $s$ computes $M=\VIRM
(\Outgoing(s),\Capacity)^{-1}\eX$ and sends the $i$th row of $M$ as
the packet over $e_i$. Note that $\eX$ contains a known ``header'',
say the $C\times C$ identity matrix over $\Field_q$, to
indicate the network transform to the receiver.

{\it Network encoders:} Let $\Outgoing (v)=\{e_1,e_2,...,e_{d}\}$ be
the outgoing edges of node $v$. For an incoming edge $e$ of $v$, $v$
computes ${\bf b}(e)=\VIRM (\Outgoing(v),d)^{-1}\VIRV{e,d}$. For the
coding coefficient $\beta(e,v,e_i)$ from $e$ via $v$ to $e_i$, $v$
sets $\beta(e,v,e_i)$ to be the $i$th component of $\mathbf {b}(e)$

{\it Receiver decoder:} The receiver receives
\begin{equation}
\eY=\Tran\eX, \label{eq:receive-mat}\end{equation} where $\Tran\in
\Field_q^{\Capacity\times \Capacity}$ can be indicated by the header
of $\eY$. If $\Tran$ is invertible the receiver can decode $\eX$
correctly.

Thus, similar to RLNC~\cite{RandCode0}, NRSC
can be implemented in a distributed manner given that each node knows its
local topology, {\it i.e.}, the adjacent neighbors. If an edge/node
has been added/deleted, only local adjustments are needed.

\subsection{Optimal throughput for multicast scenario}

The theorem below shows that with high
probability NRSC achieve the multicast capacity.
\begin{theorem}
With a probability at least $1-\Capacity|\Edges|^4/q$, receiver $r$
can decode $\eX$ correctly. \label{Th:MulticastCapa}
\end{theorem}
\noindent{\bf Proof}: Let $\mathcal X$ be the set of all random
variables involved, {\it i.e.}, $\mathcal X=\{\ID{u,v}, (u,v)\in
\Nodes\otimes \Nodes\}$.  By default we assume that any polynomial
mentioned in the proof has variables in $\mathcal X$.

Let $det_G=\Pi_{u\in \Nodes}det(u)$, where $det(u)$ is the
determinant of the matrix $\VIRM (\Outgoing(u),|\Outgoing(u)|)$ for
node $u\in \Nodes$. For each $u\in \Nodes$, since each component of
$\VIRM (\Outgoing(u),|\Outgoing(u)|)$ is a polynomial of degree at
most $|\Outgoing(u)|$, $det(u)$ is a polynomial of degree at most
$|\Outgoing(u)|^2$. Thus $det_G$ is a polynomial of degree at most
$\sum_{u\in \Nodes} |\Outgoing(u)|^2 \leq (\sum_{u\in \Nodes}
|\Outgoing(u)|)^2=|\Edges|^2$.

Let $T$ be the transform matrix from $s$ to $r$ defined in
Equation~(\ref{eq:receive-mat}). We claim each element of $det_GT$
is a polynomial of degree at most $|\Edges|^4$. To see this, we
first note that each component in $det(u)\VIRM
(\Outgoing(u),|\Outgoing(u)|)^{-1}$ is a polynomial of degree at
most $|\Outgoing(u)|^2-|\Outgoing(u)|$ (see Cramer's rule
in~\cite{CramerRule}). Thus in the construction of NRSC each local
coding coefficient $\beta(e,u,e')$ used by $u\in \Nodes$ is
$Poly_{(e,u,e')}/det(u)$, where $Poly_{(e,u,e')}$ is a polynomial of
degree at most $|\Outgoing(u)|^2$. Each element in $\Tran$ can be
expressed as $\sum_{\alpha}\bar \beta (\alpha)$, where $\bar
\beta(\alpha)=\Pi_{(e,u,e')\in \alpha} \beta(e,u,e')$ and $\alpha$
is a path from $s$ to $r$ (see~\cite{RandCode0} for references).
Thus each element in $\Tran$ can be expressed as
$Poly_\alpha/(\Pi_{u\in \alpha}det(u))$, where
$Poly_\alpha=\Pi_{(e,u,e')\in \alpha} Poly_{(e,u,e')}$. Thus
$Poly_\alpha$ is a polynomial of degree at most $\sum_{u\in \alpha}
|\Outgoing(u)|^2\leq \sum_{u\in \Nodes} |\Outgoing(u)|^2
\leq|\Edges|^2$. Since no node appears twice in a path of an acyclic
network, $det_G$ is divisible by $\Pi_{u\in \alpha}det(u)$ for each
path $\alpha$. Thus $det_G \sum_{\alpha}Poly_\alpha(\mathcal
X)/(\Pi_{u\in \alpha}det(u))$ is a polynomial of degree at most
$|\Edges|^4$. This completes the proof of the claim that each
element of $det_GT$ is a polynomial of degree at most $|\Edges|^4$.

Now we prove $det_GT$ is invertible with high probability. The
determinant of $det_GT$ is denoted as $det_r$, which is therefore a
polynomial of degree at most $|\Edges|^4\Capacity$.

Without loss of generality let
$\{\Path_1,\Path_2,...,\Path_\Capacity\}$ be the edge-disjoint paths
from the source $s$ to the receiver $r$. We first prove that $det_r$
is a nonzero polynomial, {\it i.e.}, that there exists an evaluation
of $\mathcal X$ such that $det_G\neq 0$ ({\it i.e.}, for each $u\in \Nodes$ no two edges in
$\Outgoing(u)$ have the same ID) and the
source can transmit $\Capacity$ linearly independent packets via
$\Path_1,\Path_2,...,\Path_\Capacity$.

The evaluation of $\mathcal X$ is described in the following: First,
assume each edge has a distinct ID. Second, since the $i$th outgoing
edge of the source sends the $i$th row of $M=\VIRM
(\Outgoing(s),\Capacity)^{-1}X$, the paths
$\Path_1,\Path_2,...,\Path_\Capacity$ carry linearly independent
packets on their initial edges. Third, the IDs of edges in $\Path_i$
are all changed to be the ID of the first edge in $\Path_i$. Note
that this operation preserves the property that for each $u\in \Nodes$ no two edges in
$\Outgoing(u)$ have the same ID ({\it i.e.},
$det_G\neq 0$). Finally in fact the network uses routing to transmit
the $\Capacity$ independent source packets via
$\Path_1,\Path_2,...,\Path_\Capacity$.

Thus under the above evaluation of $\mathcal X$ the matrix $det_GT$
is invertible and therefore $det_r\neq 0$. Using Schwartz-Zippel
Lemma~\cite{CCBook} $det_r\neq 0$ and thus receiver $r$ can decode
$\eX$ with a probability at least $1-|\Edges|^4\Capacity/q$ over all the evaluations of $\mathcal X$. \hfill$\Box$

Thus if the network has $k$ receivers, using the Union Bound~\cite{ProbabilityBook2005} on all receivers we conclude with a probability
at least $1-k|\Edges|^4\Capacity/q$ each receiver can decode $\eX$.

Therefor the techniques over RLNC in
multicast scenario can be directly moved into NRSC. For instance
using network error-correcting
codes~\cite{SidByAdversary}\cite{RankMetricRandCode} NRSC are able
to attain the optimal throughput for multicast with network errors.

\subsection{IRVs under NRSC}
\label{subsec:IRV-NRSC}

Following above Theorem~\ref{Th:MulticastCapa} the relations between
IRVs and network structure can be shown the same as those for
RLNC (see Lemma~\ref{LeIRVandEdges} for details). To be concrete,
for networks performing NRSC we have:

\begin{lemma}
\label{LeIRVandEdges-RS}
\begin{enumerate}
\item \label{LeCorIRVEdg-RS}The rank of the impulse response matrix $\IRM(\ErrEdg)$ of
an edge set $\ErrEdg$ with flow-rank $\SizErr $ is at most $\SizErr $.

\item \label{LeIndIRVEdg-RS} The IRVs of flow-independent edges are linear independent with a probability at least $1-\Capacity|\Edges|^4/q$.
\end{enumerate}
\end{lemma}

\noindent {\bf Proof:} The proof is similar to the proof of
Lemma~\ref{LeIRVandEdges}. \hfill $\Box$

Note that for random error model (see
Section~\ref{subsec:errormodel} for details), all tomography schemes
under RLNC are based on Lemma~\ref{LeIRVandEdges}. Thus such schemes
still work under NRSC.


\section{Locating  errors under NRSC}

\label{sec:locate-adv-rs} In this section we show that the receivers
in networks using NRSC are able to efficiently locate the network
adversaries even without the knowledge of the network topology. The
high level idea is that each column of error matrix plays the role of vector ${\bf e}$ for {\bf RS-DECODE}$(H,{\bf e})$ (see
Section~\ref{subsec:RSC} for details), where the columns of the Reed-Solomon parity-check
matrix $H$ comprise of the VIRVs of network edges. Thus the output of
algorithm {\bf RS-DECODE}$(H,{\bf v})$ locates the set of error
edges. In the end of this section, without
the priori knowledge of the network topology we provide an efficient algorithm which locates the edges suffering random errors.

{\it Assumptions and Justifications}
\begin{enumerate}
\item At most $\SizErr$ edges in $\ErrEdg$ suffer errors, {\it i.e.},
$\{e: e\in \Edges, \EdgInj(e)\neq 0\}=\ErrEdg$ and $|\ErrEdg|\leq
\SizErr$. When $2\SizErr+1\leq \Capacity$, network
error-correcting-codes (see Section~\ref{ErrorCorrectCode} for
details) are used so that the source message $\eX$ is provably
decodable .

\item Each node in $\Nodes-\{r\}$ has out-degree at least
$d=2\SizErr$. Note that Theorem~\ref{thm:high_conn_locate_adv} proves it is a necessary condition for locating $\SizErr$ errors.
\end{enumerate}
Let the elements in $\Nodes\otimes \Nodes$ be indexed by
$\{1,2,...,|\Nodes|^2\}$. The parity check matrix $H\in
\Field_q^{d\times |\Nodes|^2}$ is defined as $H=[{\bf h}_1,{\bf
h}_2,...,{\bf h}_{|\Nodes|^2}]$. Here ${\bf
h}_i$ is the VIRV (with length $d$) of the $i$th
element in $\Nodes\otimes \Nodes$. Then the adversarial error locating algorithm is:
\begin{itemize}
\item {\bf ALGORITHM VI LOCATE-ADVERSARY-RS}: The algorithm is to
locate network adversarial errors for networks performing NRSC.

\item The {\it input} of the algorithm is the source matrix $\eX$, the
parity-check matrix $H$, and the $C\times \bl$ matrix $Y$ received
by receiver $r$. The {\it output} of the algorithm is a set of edges
$\ErrEdg'$ initialized as an empty set.

\item Step A: Compute $\RevMat_{(RS,d)}=\VIRM(\Incoming(r),d)\RevMat$ and $L=\RevMat_{(RS,d)}-\eX_d$, where
$\eX_d$ comprises of the first $d$ rows of $\eX$.

\item Step B: For each column of $L$, say ${\bf v}$, compute
${\bf b}=\mbox{{\bf RS-DECODE}} (H,\mathbf v)$. If the $i$th
component of ${\bf b}$ is nonzero, the $i$th node pair $(u,v)$ in
$\Nodes\otimes \Nodes$ is added as an edge $e=(u,v)$ into
$\ErrEdg'$.

\item Step C: End {\bf LOCATE-ADVERSARY-RS}.
\end{itemize}

\begin{theorem}
The edge set $\ErrEdg'$ output by {\bf LOCATE-ADVERSARY-RS} equals actual error edge set
$\ErrEdg$. The computational complexity of {\bf LOCATE-ADVERSARY-RS}
is $\mathcal O( \bl |\Nodes|^2 d)$. \label{th:cor-LOCATE}
\end{theorem}

Before the proof we show the following key lemma when
$|\Outgoing(u)|\geq d$ for each node $u\in \Nodes-\{r\}$.
Recall that $\EdgInj(e)$ is the error packet injected on edge $e$.

\begin{lemma}
\label{thm:math-N-RSC} If the source message matrix $\eX$ equals
$0$,
\begin{equation}
\RevMat_{(RS,d)}=\sum_{e\in \Edges}\VIRV{e,d}\EdgInj(e).
\label{eq:math-N-RSC}
\end{equation}
\end{lemma}
\noindent{\bf Proof}: We proceed inductively. Throughout the proof
let $\Edges_{T}$ be the set of edges satisfying the theorem, {\it
i.e.}, $\RevMat_{(RS,d)}=\sum_{e\in \Edges}\VIRV{e,d}\EdgInj(e)$
when $\EdgInj(e)= 0$ for all $e\in \Edges-\Edges_T$.

Step A: If $\Edges_{T}=\Incoming(r)$, the theorem is true by the
definition.

Step B: Since the network is acyclic, unless $\Edges_T= \Edges$,
there must exist an edge $e\in\Edges-\Edges_T$ such that its adjacent outgoing edge
set $\Outgoing(e)$ is a subset of $\Edges_{T}$. Let
$\Outgoing(e)=\{e_1,e_2,...,e_k\}$ with $k\geq d$. If only $e$ suffers non-zero
injected errors $\EdgInj(e)$, the output of $e$ is $\EdgInj(e)$.
Thus for each $i\in[1,k]$ the output of $e_i$ is $\beta_i
\EdgInj(e)$, where $\beta_i$ is the $i$th component of ${\bf
b}(e)=\VIRM (\Outgoing(e),k)^{-1}\VIRV{e,k}$ (see
Section~\ref{subsec:NRSC-Construct} for details). Since $d\leq k$, we have $\sum_{i\in
[1,k]}\beta_i\VIRV {e_i,d}=\VIRV{e,d}$. Since $\Outgoing(e)\subseteq \Edges_T$,
$\RevMat_{(RS,d)}=\sum_{i\in[1,k]} \VIRV {e_i,d} \beta_i
\EdgInj(e)=\VIRV{e,d}\EdgInj(e)$. Therefore
Equation~(\ref{eq:math-N-RSC}) is true for the case where only $e$
suffers non-zero injected error $\EdgInj(e)$. Since NRSC are
linear codes, $e$ can be added into $\Edges_{T}$.

Step C: Since the network is acyclic and each node (or edge) in
$\Nodes$ (or $\Edges$) is connected to $r$, we can repeat
Step B until $\Edges_{T}=\Edges$. \hfill$\Box$

Recall the definition of IRV in Section~\ref{subsec:LinTran}, we
have $\RevMat=\sum_{e\in \Edges}\IRV{e}\EdgInj(e)$. Thus the
following corollary is true for network satisfying $|\Outgoing(u)|\geq d$
for each node $u\in \Nodes-\{r\}$:

\begin{corollary}
For each edge $e\in \Edges$, $\IRM(\Incoming(r),d)\IRV{e}=
\VIRV{e,d}$. \label{cor:IRVandVIRV}
\end{corollary}

For the case where no error happens in the network and the source
$s$ transmits the ${C\times n}$ message matrix $\eX$ with
$\Capacity\geq d$, by Lemma~\ref{thm:math-N-RSC} above we have
$\RevMat_{(RS,d)}=\sum_{i\in[1,\Capacity]} \VIRV {e_i,d}{\bf
x}(e_i)$, where $e_i$ is the $i$th edge of $\Outgoing(s)$ and ${\bf
x}(e_i)$ is the $i$th row of $M=\VIRM (\Outgoing(s),C)^{-1}\eX$ (see
Section~\ref{subsec:NRSC-Construct} for details). Thus $\RevMat_{(RS,d)}=\VIRM
(\Outgoing(s),d)M=\eX_d$, where $\eX_d$ is the matrix consisting of
the first $d$ rows of $\eX$.

Then we have the corollary:
\begin{corollary}
When the source message is $\eX$, $\RevMat_{(RS,d)}=\eX_d+\sum_{e\in
\Edges}\VIRV{e,d}\EdgInj(e)$. \label{cor:RevMatWithSource}
\end{corollary}

Then we can prove main theorem of this section as:

\noindent{\bf Proof of Theorem~\ref{th:cor-LOCATE}}: Using
Corollary~\ref{cor:RevMatWithSource} we have $L=\sum_{e\in \ErrEdg}
\VIRV{e,d} \EdgInj(e)$. Since $|\ErrEdg|=\SizErr\leq d/2$, each
column of $L$ is a linear combination of at most $d/2$ columns of
$H$. Additionally, since $H$ is also a parity check matrix of a
Reed-Solomon code, {\bf RS-DECODE} correctly finds all the edges
with nonzero injected errors, and therefore $\ErrEdg'=\ErrEdg$. For
each column of $L$, {\bf RS-DECODE} runs in time $\mathcal
O(|\Nodes|^2 d)$. Thus the overall time complexity of the algorithm
is $\mathcal O(\bl |\Nodes|^2 d)$. \hfill$\Box$

In the end of this section, under the condition that $|\Outgoing(u)|\geq d$ for each node $u\in \Nodes-\{r\}$, we show
that NRSC enables the receiver $r$ locate any $\SizErr\leq d-1$ random errors without the priori knowledge of the network topology.
The scheme is in the following:

\noindent{\bf Locate random errors under NRSC}: Once matrix $L$ is computed by Step A of {\bf LOCATE-ADVERSARY-RS},
for each $(u,v)\in \Nodes\otimes \Nodes$ check whether $\VIRV {(u,v),d}$ is in ${\bf L}$ ({\it i.e.}, the column space
of $L$). If so, $e(u,v)$ is output as an error edge. Continue the loop for another node pair in $\Nodes\times \Nodes$.

By Corollary~\ref{cor:IRVandVIRV} and Equation~(\ref{eq:net_tran_err}), $L=\VIRM(\ErrEdg,d)\Errors$, where
the rows of $\Errors$ comprise of $\{{\bf z}(e):e\in\ErrEdg\}$. By the proof of Lemma~\ref{LeRandomErr} we have
$\Errors$ has rank $|\ErrEdg|$ with high probability. Thus for each edge $e\in \ErrEdg$, $\VIRV {e}\in {\bf L}$. For
any edge $e'\not\in \ErrEdg$, since $\ID{e'}$ is different from any ID in $\{\ID{e}:e\in \ErrEdg\}$, $\VIRV {e',d}$ is linear
independent to the columns of $\VIRM (\ErrEdg,d)$. Thus $\VIRV {e',d}$ is not in ${\bf L}$.

\section{Topology estimation for network with random errors under NRSC}
\label{sec:Topo-Rand-NRSC} Under NRSC, the section provides a
lightweight topology estimation algorithm for the random error
model. The high level idea is that once a candidate IRV is collected
using {\bf Algorithm II, FIND-IRV} of Section~\ref{subsec:PTTopo},
the corresponding VIRV can be computed by
Corollary~\ref{cor:IRVandVIRV}. Using the VIRV the corresponding edge can be
detected. Thus {\bf Algorithm III, FIND-TOPO} is not involved, who requires {\bf FIND-TOPO}
recovering all IRV information.

For estimating the entire network topology, all assumptions in
Section~\ref{subsec:PTTopo} are required here except for
Assumption~\ref{ass:Weak-Commonrand}, which assumes weak type common
randomness.

Note that if the network
has strong connectivity and each edge suffers random error with non-negligible probability, the algorithm for locating random errors
shown in the end of Section~\ref{sec:locate-adv-rs} can also detect the topology. The algorithm shown below
only requires weak connectivity, {\it i.e.}, each internal node has out-degree at least $2$, as Assumption~\ref{ass:weak_connect}) in
Section~\ref{subsec:PTTopo}.

\begin{itemize}
\item {\bf ALGORITHM VII FIND-TOPO-RS}: Under NRSC, the algorithm is to
estimate the network topology in the presence of random
errors.

\item The {\it input} is $\{\RevMat(i),i\in[1,t]\}$, which are the received matrix for source generation
$\{1,2,...,t\}$. The {\it output} is $\Edges'$ which is a set of edges
initialized as an empty set.

\item Step A: For $i\in [1,t]$ compute $\ErrorsMat(i)_r$ as Step A in {\bf Algorithm II, FIND-IRV}

\item Step B: For any two of
$\{\ErrorsMat(i)_r,i\in[1,t]\}$, say
$\ErrorsMat(i)_r$ and $\ErrorsMat(j)_r$,
compute the intersection ${\bf \ErrorsMat(i)_r}\cap {\bf \ErrorsMat(j)_r}$. If the
intersection is a rank-one subspace $<{\bf h}>$, goto Step C. Otherwise,
continue the loop in the beginning of Step B.

\item Step C: Compute $h_1$ (and $h_2$) as the first (and second) component of $\VIRM(\Incoming(r),2){\bf h}$. For
any node pair $(u,v)\in \Nodes\otimes \Nodes$, if the ratio
$h_2/h_1$ equals $\ID{u,v}$, add $(u,v)$ as an
edge into $\Edges'$. Go back to continue the loop in the beginning of Step B.

\item Step D: End {\bf FIND-TOPO-RS}.
\end{itemize}

Let $\Graph=(\Nodes,\Edges)$ be the actual network topology, $\ProEdgeCorr$ be the probability defined
in Assumption~\ref{ass:ind_fail}) of Section~\ref{subsec:PTTopo}, $\ProRandomNonSpan$ be
$1-(1-{\SizErr}/{q})[1-2{\Capacity^2}/(\bl-\Capacity)]$ and $\ProNonAcp'$ be
$\ProEdgeCorr+2\ProRandomNonSpan+{\Capacity|\Edges|^4}/{q}$.
Then the
theorem is:
\begin{theorem}
\label{Th:Lo-Random-RS}
\begin{enumerate}
\item With a probability at most $|\Nodes|^2t^2/q$, $\Edges'$ has an edge which is not in $\Edges$.

\item If edge $e\in \ErrEdg(i)\cap\ErrEdg(j)$, $e\in
\Edges'$ with a probability at least $1-\ProNonAcp'$.
\end{enumerate}
\end{theorem}

\noindent{\bf Proof}:
\begin{enumerate}
\item Consider node pair $(u,v) \in \Nodes\otimes \Nodes$ which is not in $\Edges$.
Since $\ID{u,v}$ is independent from the network coding
 coefficients used in $\Graph$ and the random errors in each source generation, for any invocation of Step C
 the ratio $h_2/h_1$ is independent from
 $\ID{u,v}$. Thus $h_2/h_1=\ID{u,v}$ with
 probability at most $1/q$. Since there are at most $t^2$
 invocations of Step C, using Union Bound~\cite{ProbabilityBook2005} edge
 $e(u,v)$ is accepted in $\Edges'$ with a probability at most
 $t^2/q$. Since there are at most $|\Nodes|^2$ node pairs, also by
 Union Bound~\cite{ProbabilityBook2005} $\Edges'$ has an edge which is not in
 $\Edges$ with a probability at most $|\Nodes|^2t^2/q$.

\item If $e\in \ErrEdg(i)\cap\ErrEdg(j)$, from the proof of
Theorem~\ref{LeNodeTranAcc} the intersection of ${\bf
\ErrorsMat(i)_r}\cap {\bf \ErrorsMat(j)_r}$ equals $<\IRV{e}>$ with a probability at least
$1-\ProNonAcp'$. Note that the difference between $\ProNonAcp$ and $\ProNonAcp'$ comes from the difference between Lemma~\ref{LeIRVandEdges} (which is for RLNC)
and Lemma~\ref{LeIRVandEdges-RS} (which is for NRSC). Since each internal node has out-degree at least $2$, from
Corollary~\ref{cor:IRVandVIRV} we have
$\VIRM(\Incoming(r),2)\IRV{e}=\VIRV{e,2}=[\ID{e},(\ID{e})^2]^T$. It completes the
proof.
\end{enumerate}
\hfill $\Box$

\noindent{\bf Remark 1}: For estimating the failing topology ({\it i.e.},
detecting the edges with errors), even Assumption~\ref{ass:all-fail}) of Section~\ref{subsec:PTTopo}
is not needed anymore, which requires each edge suffers random
errors with a non-negligible probability. Once an edge $e$ has random errors for multiple source generations, it can be detected with high probability.

\noindent{\bf Remark 2}: For the scenario while network edges (or nodes) suffer dynamic updating, {\bf FIND-TOPO-RS} is more
robust than the topology estimation algorithm under RLNC (see Section~\ref{subsec:PTTopo} for details). The reason is that under
RLNC the receiver must use algorithm {\bf FIND-IRV} to recover all IRV information before proceed the topology estimation algorithm {\bf FIND-TOPO}.
Thus it requires the network unchanged during $t=\Theta(\log(|\Edges|)|\Edges|)$ source generations (see Remark 2 after
Theorem~\ref{LeNodeTranAcc} for details). Under NRSC,  for detecting edge $e$ {\bf FIND-TOPO-RS} only requires the network unchanged between
two fails of $e$.